\documentclass[11pt]{article}

\usepackage[a4paper,margin=1in]{geometry}
\usepackage[T1]{fontenc}
\input{glyphtounicode}
\usepackage[utf8]{inputenc}
\usepackage{lmodern}
\usepackage{amsmath}
\usepackage{graphicx}
\usepackage{booktabs}
\usepackage{tabularx}
\usepackage{array}
\usepackage{microtype}
\usepackage{placeins}
\usepackage[colorlinks=true,linkcolor=blue,citecolor=blue,urlcolor=blue]{hyperref}
\newcommand{\doi}[1]{\mbox{\scriptsize\href{https://doi.org/#1}{https://doi.org/#1}}}

\title{\textbf{A Sociophysics Approach to Measuring Democracy in the Decision Making of a Group: Three Indices}}
\author{Sudipto Roy\\
\small Department of Physics, St. Xavier's College, Kolkata, West Bengal, India.\\[-1mm]
\small email: \href{mailto:roy.sudipto@sxccal.edu}{roy.sudipto@sxccal.edu}}
\date{October 2026}

\begin{document}
\maketitle

\begin{abstract}

A group does not usually consult all its members before every decision. This paper develops a quantitative framework for measuring how widely members are consulted in decision making within a group. Two kinds of records are considered: the number of members consulted for each decision and the number of decisions in which each member is consulted. From these records, two participation indices are defined, and their equality is proved analytically. A third index measures the fraction of decisions for which at least a specified percentage of members are consulted. The distribution of consultation levels is described using a discrete maximum-entropy model from statistical physics. Tail probabilities are used to obtain bounds for the main participation index. This study examines aggregate survey data from two groups. Their observed characteristics are broadly consistent with the patterns found across the full survey. Using binomial and Poisson calculations, the future probability that all members will be consulted for a decision is examined. Inequality in consultation among members is examined through Lorenz curves, the Gini coefficient and a normalised Shannon participation measure. The member–decision structure is represented as a bipartite network. Controlled simulations show that the same overall level of consultation can have different distributions among the members. The framework measures both the overall breadth of consultation and how consultation is distributed among the members. It therefore provides a quantitative sociophysical description of one measurable aspect of democracy in decision making within a group.
\end{abstract}


\section{Introduction}

Democracy is a broad idea. It applies to countries and governments, but also to organisations and smaller groups such as committees, councils and teams. Participation in decisions that affect people is one important aspect of democracy. Democratic theory has discussed equality, participation, representation, discussion, public reasoning and the sharing of power in decision making \cite{dahl1989,pateman1970}. Different levels of participation and power have also been considered \cite{arnstein1969}, as have discussion and the exchange of reasons in collective decisions \cite{habermas1996,dryzek2000,fishkin2009}. These studies show that democracy has several dimensions. A single number cannot represent all of them.

Participation in decision making can take different forms. A person may vote, discuss an issue, give an opinion or take part in a committee. A representative may also act on behalf of others. Some systems use direct participation, while others use representative or deliberative institutions \cite{gastil2000,smith2009,geissel2018,mansbridge1999,parkinson2012}. The level of participation may also change from one decision to another. This makes it useful to study participation from the records of individual decisions within a group.

There is a large body of work on measuring democracy. Munck and Verkuilen \cite{munck2002} discussed the difficulty of defining democracy and combining its different parts into one measure. Coppedge et al. \cite{coppedge2011} stressed the importance of separate and clearly defined dimensions. The Varieties of Democracy project developed many indicators and aggregate measures \cite{lindberg2014}. Other well-known measures include the Democracy Index of the Economist Intelligence Unit \cite{eiu2025} and Polity5 \cite{marshall2020}. Some studies have also used graded measures instead of a simple yes-or-no classification \cite{elkins2000}. These studies show that a measure should clearly identify the aspect of democracy it describes.

We consider a more specific setting. A committee, department, board, team or similar group has a coordinator who makes a series of decisions. Before each decision, the coordinator may seek the views of some members. The number consulted can differ from one decision to another. We examine whether this breadth of consultation can be measured by a single number.

In this study, we define an index $I_1$ using the number of members consulted for each decision. The number consulted is divided by the number of non-coordinator members, and these fractions are averaged over decisions. Thus, one decision is the basic unit of the measure, and only the number consulted for each decision is required. We also define a member-side index, $I_2$, which is the average fraction of decisions in which each member is consulted. We prove that $I_2=I_1$ analytically. When the identities of the consulted members are available, we can also use the Gini coefficient, Lorenz curve and normalised Shannon index to study how consultation is distributed among members.

We further define the threshold index $I_3(x)$, which gives the fraction of decisions in which at least a specified percentage of members is consulted. A single threshold gives bounds on $I_1$, while the complete threshold function determines $I_1$. The distribution of the number of consulted members is described with a discrete maximum-entropy model. We also use a binomial calculation for future all-member consultation, with a Poisson approximation when the event is rare. These quantities provide different views of the same consultation records: $I_1$ gives overall breadth, $I_3(x)$ gives threshold information, and the member-wise measures describe how consultation is shared among members.

In this paper, the word ``consulted'' has a specific meaning. A member is counted as consulted when the coordinator seeks the member's opinion on a decision and gives the member an opportunity to express it. This may happen in a meeting, conversation, written request, electronic message or similar interaction. The member need not agree with the final decision, and the coordinator need not accept the opinion. Consultation is therefore different from voting or simply being present at a meeting. The same definition is used for all decisions.

The consultation measure is related to, but different from, questions studied in social choice theory. Social choice theory studies how individual preferences are combined to obtain a collective choice \cite{arrow1963,sen1970}, while voting power studies examine how much a member can affect the result of a voting rule \cite{felsenthal1998}. Studies in sociology and organisation have also considered collective action, group outcomes and participation in organisations \cite{coleman1990,march1994,cotton1988,miller1986}. The present measure instead starts from the consultation event itself: whether a member's view was sought for a particular decision.

The question is also related to sociophysics. Sociophysics uses ideas from statistical physics to study collective behaviour, social interaction and decision making \cite{macy2002,castellano2009,galam2012}. Here each consultation is treated as an interaction between a member and a decision. Probability, entropy and inequality measures are then used to study these interactions. Earlier sociophysics work has considered collective decisions and related social processes \cite{lorenz2018,chebotarev2025,marchpons2026,naumov2025,ye2026}. The present study applies this statistical approach to the breadth and distribution of consultation within groups.

\textbf{Research gap.} The literature contains many measures of democracy and many studies of participation, deliberation and collective decision making. In the literature reviewed for this study, we did not find a simple measure with the same decision-level structure as the one proposed here. This paper defines consultation breadth from the number of members consulted in each decision, proves that the member-side index gives the same value, and introduces a threshold index for the same process. We apply the framework to aggregate survey data and use maximum-entropy, inequality and network methods to examine the consultation records. Two controlled simulations show that the same overall consultation level can be distributed differently among members. The measure is intended to quantify one specific aspect of democratic decision making inside a group; it is not intended to replace broader measures of democracy. The framework can also be used to compare consultation patterns across different groups or across different periods in the same group, provided that the definition of consultation and the method of recording remain the same.

The empirical part uses confidential survey information from decision-making groups. The survey was conducted over four years through in-person interviews with coordinators and members. For the quantitative analysis, documented decision records were organised into aggregate consultation counts. These records included meeting details and minutes, consultation records, emails, written proposals and feedback, written opinions or assessments, online records and other relevant documents. Individual, group and organisational identities are not disclosed.

We use two aggregate datasets from the survey, one from a small group and one from a large group. Individual consultation histories are not available in the paper or submission package because of confidentiality. The member-wise examples used later are synthetic assignments constructed from the aggregate consultation counts; they are included to demonstrate the calculation of member-wise measures and do not represent observed individual histories.

The paper first develops the mathematical framework and describes the consultation distribution. It then applies the framework to the two aggregate datasets and proves that $I_2=I_1$. It next develops the threshold index and its bounds, considers all-member consultation probabilities, and presents the member-wise and network descriptions. The simulations then examine how different patterns of repeated consultation can change the distribution among members while keeping the overall consultation level similar. The final sections discuss the interpretation, practical requirements and limitations of the framework.
\section{The consultation index \texorpdfstring{$I_1$}{I1} and mathematical framework}

Let $m$ be the number of non-coordinator members in a group. These $m$ members form the set considered for consultation. During a chosen period, the coordinator makes $N$ decisions. We assume that this set remains fixed during the period. We treat the coordinator separately because the index measures how widely the decision process is opened to the other members.

For decision $r$, let $k_r$ be the number of members consulted according to the definition given above. Thus,
\begin{equation}
0\leq k_r\leq m,
\qquad r=1,2,\ldots,N.
\end{equation}
The value of $k_r$ gives the number of members consulted for decision $r$. It does not give their identities.

Let $n_k$ denote the number of decisions for which exactly $k$ members are consulted. Each decision has one consultation count, so the numbers $n_k$ satisfy
\begin{equation}
N=\sum_{k=0}^{m}n_k.
\label{eq:N}
\end{equation}
The fraction of decisions with consultation level $k$ is
\begin{equation}
P(k)=\frac{n_k}{N}.
\label{eq:P}
\end{equation}
Thus,
\begin{equation}
\sum_{k=0}^{m}P(k)=1.
\end{equation}
So, $P(k)$ is the probability that a randomly selected decision has exactly $k$ consulted members.

The average number of consulted members per decision is the expected value of $k$:
\begin{equation}
E=\sum_{k=0}^{m}kP(k)
=\frac{1}{N}\sum_{k=0}^{m}kn_k.
\label{eq:E}
\end{equation}
The largest possible value of $E$ is $m$, because at most $m$ members can be consulted for any decision. If all $m$ members are consulted for every decision, the average value of $k$ is $m$. We therefore use $m$ to normalise the average. The consultation participation index is defined as
\begin{equation}
I_1=\frac{E}{m}
=\frac{1}{mN}\sum_{k=0}^{m}kn_k.
\label{eq:I1}
\end{equation}

Since $P(k)=n_k/N$, we can also write the same definition as
\begin{equation}
I_1=\frac{1}{m}\sum_{k=0}^{m}kP(k)=\frac{E}{m}.
\label{eq:I1P}
\end{equation}
So, $I_1$ is the mean number of consulted members per decision divided by $m$. It is therefore the average fraction of the available members who are consulted.

The same definition can be written directly from the individual decisions, before they are grouped by $n_k$. For one decision, the fraction of available members who are consulted is $k_r/m$. Averaging this quantity over all decisions gives
\begin{equation}
I_1=\frac{1}{N}\sum_{r=1}^{N}\frac{k_r}{m}
=\frac{1}{mN}\sum_{r=1}^{N}k_r.
\label{eq:I1decision}
\end{equation}
Grouping the decisions by their value of $k_r$ changes the last sum to $\sum_{k=0}^{m}kn_k$, which gives Eq.~\eqref{eq:I1}. Thus, $I_1$ is the average of the consultation fractions for the individual decisions.

For a small example, suppose $m=4$ and there are four decisions. Let the consultation counts be $k_r=0,1,2,4$. The four decision-wise consultation fractions are $0$, $1/4$, $2/4$, and $4/4$. Their average is
\begin{equation*}
I_1=\frac{1}{4}\left(0+\frac14+\frac24+\frac44\right)=\frac{7}{16}=0.4375.
\end{equation*}
So, in this example, the average breadth of consultation is 43.75\% of the available members. It also shows why decisions with $k_r=0$ must be included when calculating $I_1$.

Equation~\eqref{eq:I1} also has a simple interpretation in terms of the records. If exactly $k$ members are consulted in a decision, that decision gives $k$ member--decision consultation events. There are $n_k$ such decisions, so they give $kn_k$ consultation events. Summing over all $k$ gives the total number of consultation events. The index is obtained by dividing this total by the largest possible number of such events, $mN$.

The bounds of the index follow directly from its definition. Since $0\leq k\leq m$ for every decision,
\begin{equation}
0\leq \frac{k}{m}\leq1.
\end{equation}
Averaging this inequality over all $N$ decisions gives
\begin{equation}
0\leq I_1\leq1.
\end{equation}
The two extreme cases have a simple meaning:
\begin{itemize}
\item $I_1=0$ if no member is consulted for any decision;
\item $I_1=1$ if all $m$ members are consulted for every decision.
\end{itemize}
Between these limits, $I_1$ gives the average fraction of the $m$ available members who are consulted. If $I_1=0.40$, for example, the average number consulted is 40 per cent of the available members. This does not mean that the same 40 per cent of members are consulted every time; the members consulted may change from one decision to another.

The index measures only the breadth of consultation. It does not measure influence on the final decision. Being consulted and having influence are not the same. For example, $I_1=0.5$ means that, on average, half of the available members are consulted; it does not mean that half of the members determine the decision. 

\subsection{Calculation of the index from actual records}

Only the number $k_r$ of consulted members is needed for each decision; the identities of the members are not required. The calculation is:
\begin{enumerate}
\item Fix the set of $m$ members and the period containing the $N$ decisions.
\item For every decision $r$, record the number $k_r$ of members who were consulted.
\item Add these numbers to obtain $\sum_r k_r$.
\item Divide the result by the maximum possible number of consultation events, $mN$.
\end{enumerate}
So, \begin{equation*}
I_1=\frac{\sum_{r=1}^{N}k_r}{mN}.
\end{equation*}
If the data are already grouped as $n_k$, we can use the equivalent formula in Eq.~\eqref{eq:I1}. Thus, the index can be calculated directly from counts. It does not require a complicated model or subjective weights for different members.

The records may be available in different levels of detail. A record containing only $k_r$ for each decision is enough to calculate $I_1$. If the identities of the consulted members are also recorded, the same data also give the member-wise quantities $F_i$, which are introduced later. These additional records also allow us to examine how consultation is distributed among members.

\subsection{Probability model for consultation levels}

We can calculate $I_1$ without fitting a probability distribution. When many decision records are available, the distribution of consultation levels provides additional information. For example, it shows whether most decisions involve a similar number of members. For decisions with at least one consulted member, we use the following discrete model as a simple description:
\begin{equation}
P_{\rm model}(k;b,c)=\frac{k^b e^{-ck}}
{\displaystyle\sum_{j=1}^{m}j^b e^{-cj}},
\qquad k=1,2,\ldots,m,
\label{eq:model}
\end{equation}
The denominator normalises the model over the finite set $1,2,\ldots,m$. In the numerical fits and examples below, we impose $b>0$ and $c>0$ as modelling restrictions; these sign restrictions do not follow from the maximum-entropy derivation alone.

\subsubsection{Maximum-entropy basis for the model}

Equation~\eqref{eq:model} is not needed to define $I_1$. We use it only to describe the distribution of consultation levels. We next show how this form follows from the maximum-entropy principle \cite{shannon1948,jaynes1957}. This principle selects the distribution with the largest Shannon entropy subject to the stated constraints. The constraint on $\ln k$ is a modelling choice that produces the factor $k^b$. We do not claim that this factor applies to all consultation data. Other probability models may be suitable for other datasets. The model does not describe how a coordinator chooses members for consultation.

Let $p_k$ denote the probability that $k$ members are consulted, with $k=1,2,\ldots,m$. The maximum-entropy calculation is a finite discrete optimisation over the probabilities $p_k$. No continuous approximation for $k$ is used in the derivation. Suppose that the following two average quantities are specified:
\begin{equation}
\sum_{k=1}^{m} k p_k=\mu
\label{eq:maxent_mean}
\end{equation}
and
\begin{equation}
\sum_{k=1}^{m} (\ln k)p_k=\ell.
\label{eq:maxent_logmean}
\end{equation}
The probabilities must also satisfy the normalisation condition
\begin{equation}
\sum_{k=1}^{m}p_k=1.
\label{eq:maxent_norm}
\end{equation}
The first condition specifies the mean number of consulted members, while the second specifies the mean of $\ln k$. Here, $\mu$ is the mean consultation level for decisions included in the model, that is, decisions with $k\geq1$. When $n_0=0$, this mean is the same as the overall mean $E=mI_1$. We maximise the Shannon entropy
\begin{equation}
H=-\sum_{k=1}^{m}p_k\ln p_k
\label{eq:entropy}
\end{equation}
subject to these three conditions.

We use the Lagrangian
\begin{align}
\mathcal{L}={}&-\sum_{k=1}^{m}p_k\ln p_k
-\lambda_0\left(\sum_{k=1}^{m}p_k-1\right)\notag\\
&-\lambda_1\left(\sum_{k=1}^{m}kp_k-\mu\right)\notag\\
&-\lambda_2\left(\sum_{k=1}^{m}(\ln k)p_k-\ell\right).
\label{eq:maxent_lagrangian}
\end{align}
At a stationary point, differentiation with respect to $p_k$ gives
\begin{equation}
-\ln p_k-1-\lambda_0-\lambda_1 k-\lambda_2\ln k=0.
\end{equation}
So, \begin{equation}
\ln p_k=C-\lambda_1 k-\lambda_2\ln k,
\end{equation}
where $C$ is a constant. Exponentiating gives
\begin{equation}
p_k=A k^b e^{-ck},
\label{eq:maxent_form}
\end{equation}
where
\begin{equation}
b=-\lambda_2,\qquad c=\lambda_1.
\end{equation}
The constant $A$ is fixed by normalisation, so
\begin{equation}
A=\left(\sum_{j=1}^{m}j^b e^{-cj}\right)^{-1}.
\label{eq:A}
\end{equation}
With this value of $A$, Eq.~\eqref{eq:maxent_form} is the same normalised distribution as Eq.~\eqref{eq:model}. $A$ is not an independent parameter of the model; once $m$, $b$ and $c$ are specified, $A$ is fixed by normalisation.

The entropy in Eq.~\eqref{eq:entropy} is strictly concave, and the three constraints are linear. If a feasible positive solution exists, this is therefore the unique maximum-entropy distribution under these constraints. This does not mean that the mean of $k$ and the mean of $\ln k$ are the only useful information about a real consultation process. They are simply the two quantities used here. The derivation gives the form of the distribution. The values of $b$ and $c$ are then estimated from the interview-based aggregates.

With the modelling restrictions $b>0$ and $c>0$, the continuous kernel has a peak at $k=b/c$ when this value lies within the consultation range.

To find the maximum of the continuous form, we consider
\begin{equation}
w(k)=k^b e^{-ck},
\qquad k>0.
\end{equation}
It is easier to differentiate its logarithm, so we write
\begin{equation}
\ln w(k)=b\ln k-ck,
\end{equation}
and hence
\begin{equation}
\frac{d}{dk}\ln w(k)=\frac{b}{k}-c.
\end{equation}
The stationary point is given by
\begin{equation}
\frac{b}{k}-c=0,
\end{equation}
which gives
\begin{equation}
k_0=\frac{b}{c}.
\label{eq:k0}
\end{equation}
The second derivative is
\begin{equation}
\frac{d^2}{dk^2}\ln w(k)=-\frac{b}{k^2}<0.
\end{equation}
This stationary point is the maximum of the continuous kernel. We use $k_0=b/c$ to locate the maximum of the continuous function. The model itself is defined only for integer values of $k$. The most probable consultation level is the integer $k\in\{1,\ldots,m\}$ for which $k^b e^{-ck}$ has the largest value. The discrete mode therefore need not be exactly equal to $k_0$.

Substituting $b=k_0c$ into Eq.~\eqref{eq:model}, the average consultation level becomes
\begin{align}
E(m,k_0,c)
&=\frac{\displaystyle\sum_{k=1}^{m}k\,k^{k_0c}e^{-ck}}
{\displaystyle\sum_{k=1}^{m}k^{k_0c}e^{-ck}}\\[1mm]
&=\frac{\displaystyle\sum_{k=1}^{m}k^{k_0c+1}e^{-ck}}
{\displaystyle\sum_{k=1}^{m}k^{k_0c}e^{-ck}}.
\end{align}
Dividing by $m$, we obtain
\begin{equation}
I_1(m,k_0,c)=
\frac{\displaystyle\sum_{k=1}^{m}k^{k_0c+1}e^{-ck}}
{\displaystyle m\sum_{k=1}^{m}k^{k_0c}e^{-ck}}.
\label{eq:I1mnc}
\end{equation}

Equation~\eqref{eq:I1mnc} gives the normalised mean of the finite discrete distribution in Eq.~\eqref{eq:model}. Thus, $I_1$ depends on $m$, $k_0$ and $c$. Here, $k_0=b/c$ locates the maximum of the continuous function, while $c$ also affects the shape of the distribution through $b=k_0c$.

\begin{figure}[htbp]
\centering
\includegraphics[width=0.88\textwidth]{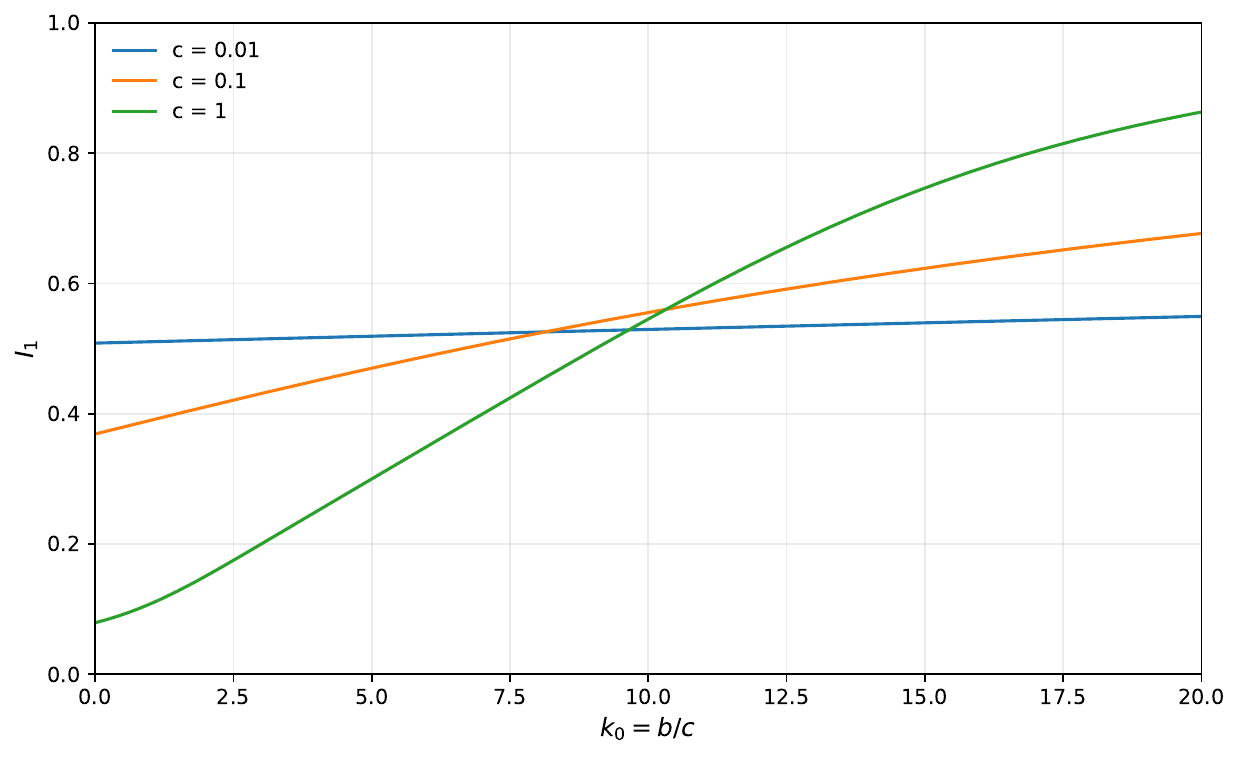}
\caption{The consultation participation index $I_1(m,k_0,c)$ as a function of $k_0=b/c$ for $m=20$ and three values of $c$, calculated from Eq.~\eqref{eq:I1mnc}. The value $k_0=0$ is shown as the limiting case $b\to0^+$; the main model assumes $b,c>0$.}
\label{fig:I1plot}
\end{figure}

We can also examine how the index changes with $k_0$, keeping $m$ and $c$ fixed. Let
\begin{equation}
a=k_0c.
\end{equation}
Then the probability in Eq.~\eqref{eq:model} is proportional to
\begin{equation}
\exp(a\ln k-ck).
\end{equation}
For this finite distribution, differentiating the normalised probability with respect to $a$ gives
\begin{equation*}
\frac{dp_k}{da}=p_k\left(\ln k-\sum_{j=1}^{m}p_j\ln j\right).
\end{equation*}
So, \begin{equation}
\frac{dE[k]}{da}
=\operatorname{Cov}(k,\ln k).
\label{eq:cov}
\end{equation}
For $m>1$, both $k$ and $\ln k$ increase with $k$. In a non-degenerate distribution, larger values of $k$ are therefore associated with larger values of $\ln k$. Hence,
\begin{equation}
\frac{dE[k]}{da}>0.
\end{equation}
Since $a=k_0c$ and $c>0$,
\begin{equation}
\frac{dE[k]}{dk_0}=c\frac{dE[k]}{da}>0.
\end{equation}
So, \begin{equation}
\boxed{\frac{\partial I_1}{\partial k_0}>0}
\end{equation}
for fixed $m$ and $c$ when $m>1$. This explains why the curves in Fig.~\ref{fig:I1plot} increase with $k_0$. This does not mean that $k_0$ alone can be used to compare two groups with different values of $c$, because $I_1$ depends on both parameters.

We define the model in Eq.~\eqref{eq:model} for $k=1,2,\ldots,m$, although the general definition of $I_1$ also allows $k=0$. A decision with $k=0$ must still be included when calculating $I_1$. We do not include $k=0$ in the maximum-entropy fit because the model uses $\ln k$, which is not defined there. Therefore, we fit the conditional distribution of decisions with $k\geq1$. For the large group, we leave the two decisions with $k=0$ out of the fitting denominator. They remain in the full empirical distribution and in the calculation of $I_1$. A separate probability for $k=0$ can be added if a model for the complete distribution is required.

For illustration, Fig.~\ref{fig:I1plot} shows $I_1(m,k_0,c)$ from Eq.~\eqref{eq:I1mnc} for $m=20$, with $k_0$ varied from 0 to 20 for three fixed values of $c$. The value $k_0=0$ is shown as the limiting case $b\to0^+$. For the main model, $b,c>0$ and hence $k_0>0$.

\FloatBarrier

\subsection{Cumulative consultation distribution}

The average value $I_1$ gives one overall number, but it does not tell us how often a particular consultation level is reached. To describe this, let
\begin{equation}
N_{\geq k}=\sum_{j=k}^{m}n_j
\label{eq:Ngeq}
\end{equation}
be the number of decisions involving at least $k$ members. The corresponding cumulative probability is
\begin{equation}
P_{\geq}(k)=\frac{N_{\geq k}}{N}
=\sum_{j=k}^{m}P(j).
\label{eq:cumulative}
\end{equation}
For example, $P_{\geq}(m)$ is the fraction of decisions with all-member consultation, while $P_{\geq}(\lceil m/2\rceil)$ is the fraction of decisions in which at least half of the members are consulted. These quantities provide information that $I_1$ alone does not give.

\section{Data and analysis}

\subsection{Data collection methodology}

We collected the survey information over four years through confidential, in-person interviews with coordinators and members of decision-making groups. The survey covered 102 groups: 48 small groups and 54 large groups. The groups included committees, academic departments, associations, societies and similar bodies. Each selected group had a coordinator formally assigned by the organisation, whose role included making decisions in consultation with other members.

We classify the groups by the observed value of $n_0$, the number of decisions in which no member was consulted. Groups with $n_0=0$ are called 'small groups' and those with $n_0>0$ are called 'large groups'. In the surveyed data, the 48 small groups had $m\leq30$ and $n_0=0$, while the 54 large groups had $m>30$ and $n_0>0$. Thus, $m=30$ is a boundary observed in this sample, not a general rule for classifying groups. The large-group example in Table~\ref{tab:table2} has $n_0=2$.

The survey data for all 102 groups showed the same broad pattern: the $n_k$ values generally rose to a peak and then decreased as $k$ increased, while decisions involving about 80\% or more of the members were relatively few. These observations provide the wider context for the two groups analysed in detail.

The interviews collected information on group size, decisions or meetings, the issues considered and the extent to which members' opinions were sought. Respondents also provided information on decision events during the preceding five years, ending on the interview date. The numerical analysis uses documented decision records obtained during the interviews. These included meeting details and minutes, consultation records, emails, written proposals and feedback, written opinions or assessments, online records, personal written records or evaluations, signed decisions or resolutions, decision proposals and other relevant documents. We reviewed these records and used them to prepare the aggregate consultation counts for the analysis. The records were used to measure consultation breadth, not to identify or assess any person or organisation.

Tables~\ref{tab:table1} and~\ref{tab:table2} give group-level aggregates in the form of consultation-count frequencies $n_k$. They do not give individual consultation histories. Tables~\ref{tab:table3} and~\ref{tab:table4} use a nested assignment to construct illustrative member-wise frequencies from the aggregate data. These assignments are synthetic and do not represent observed member-level histories.

The definition of consultation given in Section~1 was used throughout the data collection. Simply being a formal member, being present or being entitled to participate did not count as consultation. The coordinator had to seek the member's view and give the member an opportunity to express it. The member did not have to agree with the final decision, and the coordinator did not have to accept or follow the view.

The small-group aggregate used in the analysis has $m=22$ and covers $N=225$ decision events. The large-group aggregate has $m=105$ and covers $N=1062$ decision events. These two case groups were selected from the survey because their observed characteristics are broadly consistent with the corresponding group types. They are not statistical estimates for all 48 small groups or all 54 large groups. The full survey provides the wider context for these two cases. The two five-year periods end on their respective interview dates.

\subsection{Small-group aggregate}

\begin{figure}[htbp]
\centering
\includegraphics[width=0.88\textwidth]{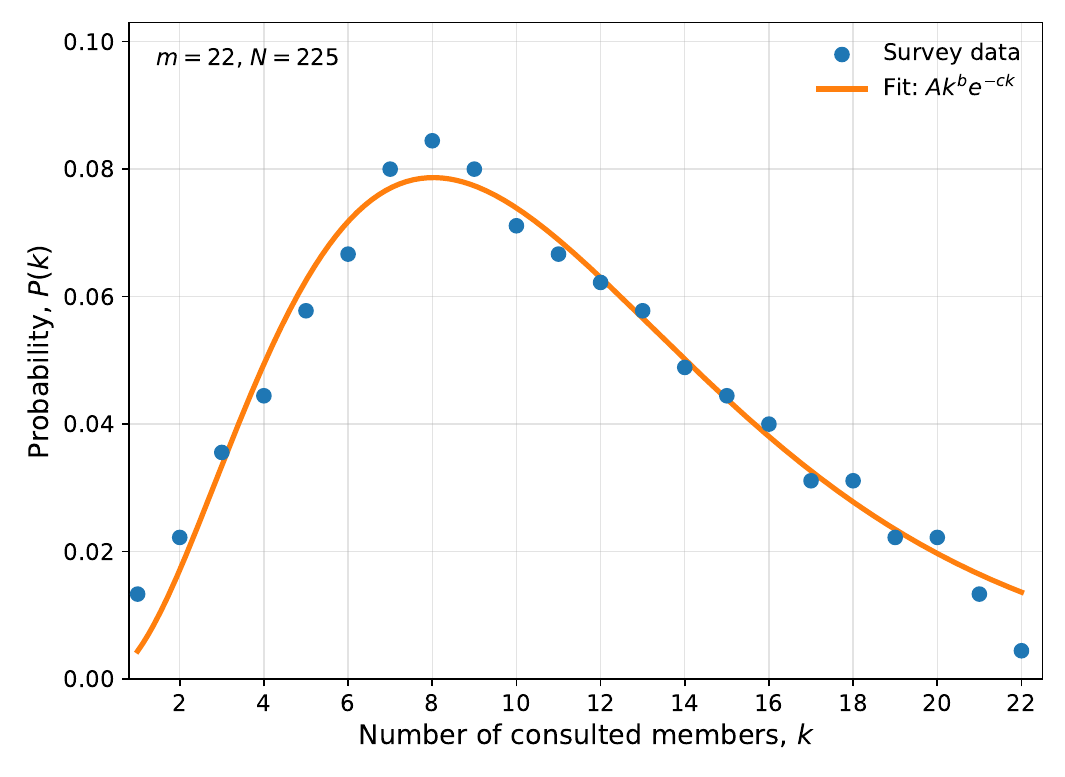}
\caption{Aggregate consultation distribution obtained from the confidential survey of the small group in Table~\ref{tab:table1}, together with its fit to Eq.~\eqref{eq:model}. The plotted points represent the group-level aggregate. The fit uses the conditional distribution for $k\geq1$.}
\label{fig:smallfit}
\end{figure}

We first consider the surveyed small group with $m=22$ and $N=225$ decision events. Table~\ref{tab:table1} gives the values of $n_k$ for $k=0,1,\ldots,22$. Here $n_0=0$. The counts rise gradually to a maximum at $k=8$. They then decrease. Figure~\ref{fig:smallfit} shows the consultation distribution obtained from the survey information. We did not construct this distribution from the fitting formula.

For the fit, we minimise the sum of squared residuals numerically over $b>0$ and $c>0$:
\begin{equation}
S(b,c)=\sum_{k=1}^m\left[P_{\rm data}(k)-P_{\rm model}(k;b,c)\right]^2,
\label{eq:S}
\end{equation}
where $P_{\rm model}(k;b,c)$ is given by Eq.~\eqref{eq:model}. For the small-group data, $n_0=0$, so $P_{\rm data}(k)=n_k/N$. For the large-group data, $n_0>0$. The fit therefore uses the conditional empirical probability $P_{\rm data}(k)=n_k/(N-n_0)$ for $k\geq1$. The same fitting criterion is used for both datasets.

We use the usual form for $R^2$
\begin{equation}
R^2=1-\frac{\sum_{k=1}^{m}[P_{\rm data}(k)-P_{\rm model}(k)]^2}{\sum_{k=1}^{m}[P_{\rm data}(k)-\overline{P}_{\rm data}]^2},
\label{eq:R2}
\end{equation}
where $\overline{P}_{\rm data}$ is the mean of the empirical probabilities over the fitted values $k=1,\ldots,m$. For the large group, these empirical probabilities are conditional on $k\geq1$, because the two $k=0$ decisions are not included in the fitted distribution. Thus, $R^2$ describes how closely the fitted conditional probabilities follow the empirical probabilities used in the fit. It is not a goodness-of-fit measure for the full distribution including $k=0$. We use it only as a simple measure of agreement between the fitted and observed distributions.

\begin{table}[htbp]
\centering
\caption{Aggregated consultation data from the surveyed small group with $m=22$ and $N=225$ decision events. Here $n_0=0$. The data rise gradually to a maximum at $k=8$ and then decrease. The displayed probabilities are rounded to four decimal places.}
\label{tab:table1}
\small
\begin{tabular}{rrrrrr}
\toprule
$k$ & $n_k$ & $P(k)$ & $kP(k)$ & $N_{\geq k}$ & $P_{\geq}(k)$\\
\midrule
0 & 0 & 0.0000 & 0.0000 & 225 & 1.0000\\
1 & 3 & 0.0133 & 0.0133 & 225 & 1.0000\\
2 & 5 & 0.0222 & 0.0444 & 222 & 0.9867\\
3 & 8 & 0.0356 & 0.1067 & 217 & 0.9644\\
4 & 10 & 0.0444 & 0.1778 & 209 & 0.9289\\
5 & 13 & 0.0578 & 0.2889 & 199 & 0.8844\\
6 & 15 & 0.0667 & 0.4000 & 186 & 0.8267\\
7 & 18 & 0.0800 & 0.5600 & 171 & 0.7600\\
8 & 19 & 0.0844 & 0.6756 & 153 & 0.6800\\
9 & 18 & 0.0800 & 0.7200 & 134 & 0.5956\\
10 & 16 & 0.0711 & 0.7111 & 116 & 0.5156\\
11 & 15 & 0.0667 & 0.7333 & 100 & 0.4444\\
12 & 14 & 0.0622 & 0.7467 & 85 & 0.3778\\
13 & 13 & 0.0578 & 0.7511 & 71 & 0.3156\\
14 & 11 & 0.0489 & 0.6844 & 58 & 0.2578\\
15 & 10 & 0.0444 & 0.6667 & 47 & 0.2089\\
16 & 9 & 0.0400 & 0.6400 & 37 & 0.1644\\
17 & 7 & 0.0311 & 0.5289 & 28 & 0.1244\\
18 & 7 & 0.0311 & 0.5600 & 21 & 0.0933\\
19 & 5 & 0.0222 & 0.4222 & 14 & 0.0622\\
20 & 5 & 0.0222 & 0.4444 & 9 & 0.0400\\
21 & 3 & 0.0133 & 0.2800 & 4 & 0.0178\\
22 & 1 & 0.0044 & 0.0978 & 1 & 0.0044\\
\bottomrule
\end{tabular}
\end{table}

From the table, the total number of member--decision consultation events is
\begin{equation*}
\sum_{k=0}^{22}kn_k=2307.
\end{equation*}
So, \begin{equation*}
E=\frac{2307}{225}=10.253333,
\end{equation*}
and
\begin{equation*}
I_1=\frac{2307}{22\times225}=0.4660606\ldots\approx0.4661.
\end{equation*}
On average, 10.253333 members are consulted per decision. This is about 46.61 per cent of the available members. The average does not mean that the same members are consulted for every decision.

There are $N_{\geq 12}=85$ decisions in which more than half of the 22 members are consulted. Therefore,
\begin{equation*}
P_{\geq}(12)=\frac{85}{225}=0.3778.
\end{equation*}
So, 37.78\% of the decisions in this aggregate involve more than half of the available members.

For Table~\ref{tab:table1}, the fitted values are
\begin{equation*}
b\approx2.3895757,\qquad c\approx0.29784146,
\end{equation*}
and so
\begin{equation*}
A\approx0.00592391.
\end{equation*}
The sum of squared residuals is about $0.000369$. The coefficient of determination is
\begin{equation*}
R^2\approx0.9692.
\end{equation*}
The fit is reasonably good, although the aggregate data still show some visible deviations from the fitted curve. The counts came from the survey information; they were not generated from Eq.~\eqref{eq:model}. The fitted curve is used only as a mathematical description of the observed consultation distribution.

For the fitted values, Eq.~\eqref{eq:k0} gives
\begin{equation*}
k_0=\frac{b}{c}\approx8.0230.
\end{equation*}
Comparing the allowed integer values shows that the discrete distribution is largest at $k=8$. The quantity $b/c$ gives the location of the maximum of the continuous form. The most probable number of consulted members is found by comparing the probabilities at the allowed integer values.

\subsection{Large-group aggregate}

\begin{figure}[htbp]
\centering
\includegraphics[width=0.88\textwidth]{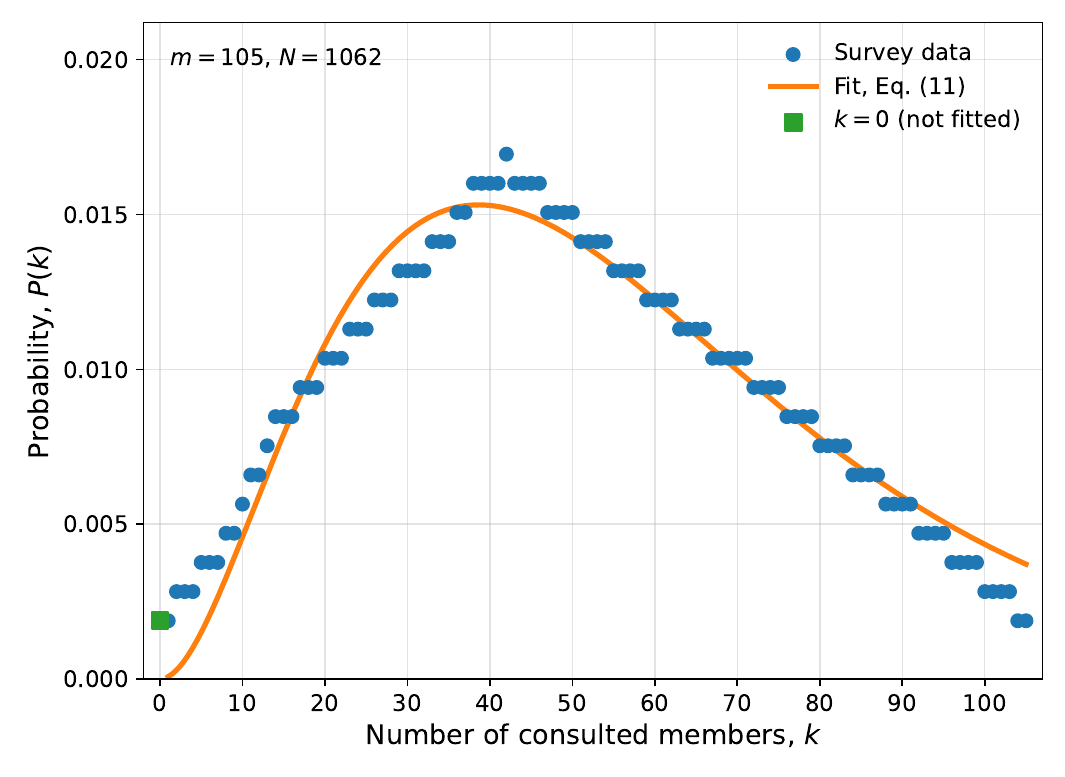}
\caption{Aggregate consultation distribution obtained from the confidential survey of the large group in Table~\ref{tab:table2}, together with its fit to Eq.~\eqref{eq:model}. The plotted points represent the group-level aggregate. We fit the conditional distribution for $k\geq1$; the two $k=0$ decisions are not fitted. For plotting against the full empirical distribution, the fitted conditional probabilities are multiplied by $(N-n_0)/N$.}
\label{fig:largefit}
\end{figure}

The second dataset comes from the surveyed large group. It has $m=105$ and $N=1062$ decision events. Table~\ref{tab:table2} gives the consultation distribution. Here $n_0=2$. Therefore, two decision events involved no member. The distribution rises gradually to a maximum at $k=42$. It then decreases. Figure~\ref{fig:largefit} shows the aggregate consultation pattern of this surveyed large group.

\begin{table}[htbp]
\centering
\caption{Aggregated consultation data from the surveyed large group with $m=105$ and $N=1062$ decision events. Here $n_0=2$. The data rise gradually to a maximum at $k=42$ and then decrease.}
\label{tab:table2}
\scriptsize
\setlength{\tabcolsep}{3.5pt}
\begin{tabular}{r r r @{\qquad} r r r}
\toprule
$k$ & $n_k$ & $P(k)$ & $k$ & $n_k$ & $P(k)$\\
\midrule
0 & 2 & 0.0019 & 53 & 15 & 0.0141\\
1 & 2 & 0.0019 & 54 & 15 & 0.0141\\
2 & 3 & 0.0028 & 55 & 14 & 0.0132\\
3 & 3 & 0.0028 & 56 & 14 & 0.0132\\
4 & 3 & 0.0028 & 57 & 14 & 0.0132\\
5 & 4 & 0.0038 & 58 & 14 & 0.0132\\
6 & 4 & 0.0038 & 59 & 13 & 0.0122\\
7 & 4 & 0.0038 & 60 & 13 & 0.0122\\
8 & 5 & 0.0047 & 61 & 13 & 0.0122\\
9 & 5 & 0.0047 & 62 & 13 & 0.0122\\
10 & 6 & 0.0056 & 63 & 12 & 0.0113\\
11 & 7 & 0.0066 & 64 & 12 & 0.0113\\
12 & 7 & 0.0066 & 65 & 12 & 0.0113\\
13 & 8 & 0.0075 & 66 & 12 & 0.0113\\
14 & 9 & 0.0085 & 67 & 11 & 0.0104\\
15 & 9 & 0.0085 & 68 & 11 & 0.0104\\
16 & 9 & 0.0085 & 69 & 11 & 0.0104\\
17 & 10 & 0.0094 & 70 & 11 & 0.0104\\
18 & 10 & 0.0094 & 71 & 11 & 0.0104\\
19 & 10 & 0.0094 & 72 & 10 & 0.0094\\
20 & 11 & 0.0104 & 73 & 10 & 0.0094\\
21 & 11 & 0.0104 & 74 & 10 & 0.0094\\
22 & 11 & 0.0104 & 75 & 10 & 0.0094\\
23 & 12 & 0.0113 & 76 & 9 & 0.0085\\
24 & 12 & 0.0113 & 77 & 9 & 0.0085\\
25 & 12 & 0.0113 & 78 & 9 & 0.0085\\
26 & 13 & 0.0122 & 79 & 9 & 0.0085\\
27 & 13 & 0.0122 & 80 & 8 & 0.0075\\
28 & 13 & 0.0122 & 81 & 8 & 0.0075\\
29 & 14 & 0.0132 & 82 & 8 & 0.0075\\
30 & 14 & 0.0132 & 83 & 8 & 0.0075\\
31 & 14 & 0.0132 & 84 & 7 & 0.0066\\
32 & 14 & 0.0132 & 85 & 7 & 0.0066\\
33 & 15 & 0.0141 & 86 & 7 & 0.0066\\
34 & 15 & 0.0141 & 87 & 7 & 0.0066\\
35 & 15 & 0.0141 & 88 & 6 & 0.0056\\
36 & 16 & 0.0151 & 89 & 6 & 0.0056\\
37 & 16 & 0.0151 & 90 & 6 & 0.0056\\
38 & 17 & 0.0160 & 91 & 6 & 0.0056\\
39 & 17 & 0.0160 & 92 & 5 & 0.0047\\
40 & 17 & 0.0160 & 93 & 5 & 0.0047\\
41 & 17 & 0.0160 & 94 & 5 & 0.0047\\
42 & 18 & 0.0169 & 95 & 5 & 0.0047\\
43 & 17 & 0.0160 & 96 & 4 & 0.0038\\
44 & 17 & 0.0160 & 97 & 4 & 0.0038\\
45 & 17 & 0.0160 & 98 & 4 & 0.0038\\
46 & 17 & 0.0160 & 99 & 4 & 0.0038\\
47 & 16 & 0.0151 & 100 & 3 & 0.0028\\
48 & 16 & 0.0151 & 101 & 3 & 0.0028\\
49 & 16 & 0.0151 & 102 & 3 & 0.0028\\
50 & 16 & 0.0151 & 103 & 3 & 0.0028\\
51 & 15 & 0.0141 & 104 & 2 & 0.0019\\
52 & 15 & 0.0141 & 105 & 2 & 0.0019\\
\bottomrule
\end{tabular}
\end{table}

For this dataset, the total number of member--decision consultation events is
\begin{equation*}
\sum_{k=0}^{105}kn_k=52341.
\end{equation*}
So, the average number of consulted members is
\begin{equation*}
E=\frac{52341}{1062}=49.2853107,
\end{equation*}
and the consultation participation index is
\begin{equation*}
I_1=\frac{52341}{105\times1062}=0.4693839\ldots\approx0.4694.
\end{equation*}
Thus, the average consultation level is about 46.94 per cent of the available members.

The same least-squares criterion in Eq.~\eqref{eq:S} is applied to this dataset with $m=105$. The fitted parameters are
\begin{equation*}
b\approx1.9748150,\qquad c\approx0.05110933,
\end{equation*}
with
\begin{equation*}
A\approx0.00008117.
\end{equation*}
The sum of squared residuals is about $9.66\times10^{-5}$ and
\begin{equation*}
R^2\approx0.9500.
\end{equation*}
For comparison with the full data, we multiply the fitted conditional probabilities by $(N-n_0)/N$. The fit is reasonably good, but it is not exact. The fitted value $k_0=b/c\approx38.639$ lies within the allowed range $1\leq k\leq105$. Therefore, the continuous kernel has an interior maximum near $k=39$. The observed data have their maximum at $k=42$. The fitted curve gives a smooth description of the observed data, but it does not match every feature of the distribution.

\FloatBarrier

\section{Member-side consultation index \texorpdfstring{$I_2$}{I2}}

So far, we have viewed the records from the side of the decisions. For each decision, we asked how many members were consulted. We now view the same records from the side of the members. Let $F_i$ be the number of decisions for which member $i$ is consulted. We define
\begin{equation}
f_i=\frac{F_i}{N},
\end{equation}
so that $f_i$ is the fraction of decisions involving member $i$.

If every member is consulted in every decision, then $F_i=N$ for all $i$, and hence
\begin{equation}
\left[\sum_{i=1}^{m}F_i\right]_{\rm full}=mN.
\end{equation}
We define the member-based index as
\begin{equation}
I_2=\frac{\displaystyle\sum_{i=1}^{m}F_i}{mN}
=\frac{1}{m}\sum_{i=1}^{m}f_i.
\label{eq:I2}
\end{equation}
Clearly,
\begin{equation}
0\leq I_2\leq1.
\end{equation}

\subsection{\texorpdfstring{Relation between $I_1$ and $I_2$}{Relation between I1 and I2}}

We do not assume a relation between $I_1$ and $I_2$. We obtain their relation directly from the participation records by counting the same member--decision consultation events in two ways.

Label the decisions by $r=1,2,\ldots,N$. Let $\mathcal C_r$ be the set of members consulted for decision $r$. Then $k_r=|\mathcal C_r|$. Define
\begin{equation*}
x_{ir}=\begin{cases}
1,&\text{if member $i$ is consulted for decision $r$,}\\
0,&\text{otherwise.}
\end{cases}
\end{equation*}
For a fixed member $i$,
\begin{equation}
F_i=\sum_{r=1}^{N}x_{ir}.
\end{equation}
So, \begin{equation}
\sum_{i=1}^{m}F_i
=\sum_{r=1}^{N}\sum_{i=1}^{m}x_{ir}.
\end{equation}
For decision $r$, exactly $k_r$ members are consulted, so
\begin{equation}
\sum_{i=1}^{m}x_{ir}=k_r.
\end{equation}
So, \begin{equation}
\sum_{i=1}^{m}F_i=\sum_{r=1}^{N}k_r.
\end{equation}
Grouping the decisions by consultation size gives
\begin{equation}
\sum_{r=1}^{N}k_r=\sum_{k=0}^{m}kn_k.
\end{equation}
So, \begin{equation}
\boxed{\displaystyle\sum_{i=1}^{m}F_i=\sum_{k=0}^{m}kn_k.}
\label{eq:counting}
\end{equation}

Equation~\eqref{eq:counting} has a simple interpretation. We can count the consultation events member by member, which gives $\sum_i F_i$. We can also count them decision by decision, which gives $\sum_k k n_k$. Both sums count the same events, so they are equal.

Substituting Eq.~\eqref{eq:counting} into Eq.~\eqref{eq:I2},
\begin{equation*}
\begin{aligned}
I_2
&=\frac{1}{mN}\sum_{i=1}^{m}F_i\\
&=\frac{1}{mN}\sum_{k=0}^{m}kn_k\\
&=\frac{1}{m}\left(\frac{1}{N}\sum_{k=0}^{m}kn_k\right)\\
&=\frac{E}{m}=I_1.
\end{aligned}
\end{equation*}
So, \begin{equation}
\boxed{I_2=I_1.}
\label{eq:equality}
\end{equation}

This equality is useful because the two forms can be calculated from different types of records. One uses the number of members consulted for each decision; the other uses the number of decisions in which each member was consulted. When the two records describe the same set of consultation events and decisions, the overall index is exactly the same.

\subsection{Usefulness of the member-side index}\label{sec:usefulness_dprime}

A researcher may not have records for each decision. Instead, the available records may give the fraction of decisions in which each member's opinion was sought. Averaging these fractions over the $m$ members gives $I_2$. So, the member-side index can be found without recording the consultation count for every decision. The same definition of consultation and the same period should be used for all members. Since $I_2=I_1$, this member-side method gives the same overall index when the records describe the same set of decisions and consultation events.

\section{\texorpdfstring{Threshold-based consultation index ($I_3(x)$)}{Threshold-based consultation index (I-3(x))}}

The index $I_1$ gives the average fraction of members consulted. A threshold answers a different question directly. For example, an organisation may specify that a decision should involve at least 70\% of the members. The threshold index gives the fraction of decisions that reach this level.

Let $x$ be the threshold percentage chosen by the organisation, with $0<x\leq100$. The smallest whole number of members corresponding to this threshold is
\begin{equation}
q_x=\left\lceil\frac{xm}{100}\right\rceil.
\label{eq:qx}
\end{equation}
where $\lceil\cdot\rceil$ means that the value is rounded up to the next whole number when necessary. We define the threshold-based consultation index as
\begin{equation}
I_3(x)=\frac{1}{N}\sum_{r=1}^{N} I(k_r\geq q_x).
\label{eq:I3}
\end{equation}
Here, $I(k_r\geq q_x)$ is 1 when at least $q_x$ members are consulted for decision $r$ and 0 otherwise. Therefore,
\begin{equation}
I_3(x)=\frac{1}{N}\sum_{k=q_x}^{m}n_k.
\label{eq:I3_nk}
\end{equation}
So, $I_3(x)$ is the fraction of decisions in which at least $x\%$ of the members are consulted. It is also the cumulative probability
\begin{equation}
I_3(x)=P_{\geq}(q_x).
\label{eq:I3_PC}
\end{equation}

\begin{center}
\fbox{%
\begin{minipage}{0.88\textwidth}
\textbf{Three indices at a glance.}

\textbf{$I_1$} --- The average fraction of the members whose opinions are sought for each decision.

\textbf{$I_2$} --- The average fraction of decisions in which a member's opinion is sought. This paper proves that $I_2=I_1$.

\textbf{$I_3(x)$} --- The fraction of decisions in which the opinions of at least $x\%$ of the members are sought.
\end{minipage}%
}
\end{center}
\subsection{\texorpdfstring{Exact relation between $I_1$ and $I_3(x)$}{Exact relation between I1 and I3(x)}}

The complete threshold function $I_3(x)$ contains the same consultation-level information as the distribution of the number of consulted members. This gives an exact relation for $I_1$. Let $K$ denote the number of consulted members for a randomly selected decision. Define
\begin{equation}
T_j=P(K\geq j),\qquad j=1,2,\ldots,m.
\label{eq:Tj}
\end{equation}
This is the standard tail-sum identity for a non-negative integer-valued consultation count. The full set of tail probabilities therefore contains the same information about the mean as the full distribution. From the definition of $I_3(x)$, choosing the threshold $x=100j/m$ gives
\begin{equation}
T_j=I_3\!\left(\frac{100j}{m}\right).
\label{eq:TjI3}
\end{equation}
For the non-negative integer-valued variable $K$, the tail-sum identity gives
\begin{equation}
E[K]=\sum_{j=1}^{m}P(K\geq j)=\sum_{j=1}^{m}T_j.
\label{eq:tailsum}
\end{equation}
Since $I_1=E[K]/m$, Eqs.~\eqref{eq:TjI3} and~\eqref{eq:tailsum} give
\begin{equation}
\boxed{\displaystyle
I_1=\frac{1}{m}\sum_{j=1}^{m}I_3\!\left(\frac{100j}{m}\right).}
\label{eq:I1_I3_sum}
\end{equation}
So, $I_1$ is the average of the threshold values of $I_3(x)$ at all attainable integer consultation levels.

The same result can also be written as an integral. Since $I_3(x)$ is a step function based on the integer-valued consultation count, it is constant over each interval $\left((j-1)100/m,\,j100/m\right]$. Hence,
\begin{equation}
\boxed{\displaystyle
I_1=\frac{1}{100}\int_{0}^{100}I_3(x)\,dx.}
\label{eq:I1_I3_integral}
\end{equation}
The value at the single point $x=0$ does not affect the integral. So, the complete $I_3(x)$ curve determines $I_1$ exactly, while an individual threshold value $I_3(x)$ gives additional information about how often that consultation level is reached.

The measures $I_1$ and $I_3(x)$ answer related but different questions. $I_1$ gives the average breadth of consultation, while $I_3(x)$ shows how often a chosen minimum level is reached. The threshold $x$ can be chosen according to the rule or expectation of the organisation. A single threshold value does not replace $I_1$, but the complete threshold function contains enough information to recover $I_1$ exactly.

\subsection{\texorpdfstring{Bounds on $I_1$ from a single threshold}{Bounds on I1 from a single threshold}}

The exact relation above uses the full set of threshold values. In practice, an organisation may record only whether a chosen threshold was reached. We then need to determine what can still be said about $I_1$. Let
\begin{equation}
\tau_x=I_3(x),\qquad q_x=\left\lceil\frac{xm}{100}\right\rceil.
\label{eq:pix_threshold}
\end{equation}
Here $\tau_x$ is the fraction of decisions in which at least $q_x$ members were consulted. In those decisions, the number consulted is at least $q_x$. In the remaining fraction $1-\tau_x$ of decisions, the number consulted is at most $q_x-1$. Therefore, for the consultation count $K$,
\begin{equation}
q_x\tau_x\leq E[K]\leq m\tau_x+(q_x-1)(1-\tau_x).
\label{eq:EK_bounds}
\end{equation}
The lower bound occurs when the decisions that reach the threshold involve exactly $q_x$ members and all other decisions involve no members. The upper bound occurs when the decisions that reach the threshold involve all $m$ members and all other decisions involve exactly $q_x-1$ members. These two arrangements show that the bounds are sharp.

Since $I_1=E[K]/m$, Eq.~\eqref{eq:EK_bounds} gives
\begin{equation}
\boxed{\displaystyle
\frac{q_x}{m}I_3(x)\leq I_1\leq I_3(x)+\frac{q_x-1}{m}\left[1-I_3(x)\right].}
\label{eq:I1_bounds_I3}
\end{equation}
So, even one threshold value gives a justified interval for the average consultation index. The interval becomes exact only when the complete threshold function is known, as shown above. This result is useful when an organisation can reliably record whether a chosen minimum level of consultation was reached. It is especially useful when the organisation does not record the exact number of members consulted for every decision.

The 50\% threshold in Table~\ref{tab:threshold} gives a direct illustration. For the small group, $m=22$, $q_{50}=11$, and $I_3(50\%)=100/225=0.4444$. Equation~\eqref{eq:I1_bounds_I3} gives
\begin{equation}
0.2222\leq I_1\leq0.6970,
\end{equation}
while the actual value obtained from the complete aggregate data is $I_1=0.4661$. For the large group, $m=105$, $q_{50}=53$, and $I_3(50\%)=453/1062=0.4266$. The corresponding bounds are
\begin{equation}
0.2153\leq I_1\leq0.7105,
\end{equation}
while the actual value is $I_1=0.4694$. These examples show why one threshold is useful, but not enough to determine the average consultation exactly.

For the two surveyed groups, the values are shown in Table~\ref{tab:threshold}. The threshold is converted into a whole number of members using Eq.~\eqref{eq:qx}.

\begin{table}[htbp]
\centering
\caption{Threshold-based consultation index $I_3(x)$ for the two surveyed groups. $q_x$ is the minimum whole number of members that must be consulted to reach the threshold $x$.}
\label{tab:threshold}
\small
\begin{tabular}{c c c c c c}
\toprule
Threshold $x$ & \multicolumn{2}{c}{Small group} & \multicolumn{2}{c}{Large group}\\
\cmidrule(lr){2-3}\cmidrule(lr){4-5}
 & $q_x$ & $I_3(x)$ & $q_x$ & $I_3(x)$\\
\midrule
40\% & 9  & $134/225=0.5956$ & 42 & $633/1062=0.5960$\\
50\% & 11 & $100/225=0.4444$ & 53 & $453/1062=0.4266$\\
60\% & 14 & $58/225=0.2578$  & 63 & $315/1062=0.2966$\\
70\% & 16 & $37/225=0.1644$  & 74 & $192/1062=0.1808$\\
80\% & 18 & $21/225=0.0933$  & 84 & $104/1062=0.0979$\\
\bottomrule
\end{tabular}
\end{table}

At the 70\% threshold, $I_3(70\%)=0.1644$ for the small group. So, 16.44\% of its decisions involved at least 16 of the 22 members. For the large group, $I_3(70\%)=0.1808$, so 18.08\% of its decisions involved at least 74 of the 105 members. These values also illustrate why one threshold cannot determine $I_1$. The corresponding bounds are $0.1196\leq I_1\leq0.7341$ for the small group and $0.1274\leq I_1\leq0.7503$ for the large group. The intervals are wide because the threshold does not specify the exact consultation counts below it. As the threshold increases, $I_3(x)$ can only decrease or remain unchanged.

The threshold index is useful because an average can hide how consultation is spread across decisions. Two groups may have similar values of $I_1$ but differ in how often many members are consulted. $I_3(x)$ shows this part of the pattern directly. Like $I_1$, it measures consultation only; it does not show whether the opinions were followed in decisions taken or whether members had equal influence on the final decision.

\section{Binomial model for all-member consultation}

\subsection{All-member consultation rate}

At the 100 per cent threshold, $k=m$. We call this \emph{all-member consultation} because every one of the $m$ non-coordinator members has been consulted. It is the broadest level of consultation that can be identified from these data. It is not a complete measure of democracy because it does not show how much influence members have on the final decision. It measures participation through consultation before a decision is made.

The observed fraction of decisions with all-member consultation is
\begin{equation}
p_{\rm all}=P(m)=\frac{n_m}{N}.
\label{eq:q}
\end{equation}
Here $p_{\rm all}$ is the observed probability of a decision with all-member consultation. For each decision, let $I(k_r=m)$ be 1 when all $m$ members are consulted and 0 otherwise. Since $I(k_r=m)\leq k_r/m$ for every decision, averaging over all decisions gives
\begin{equation}
p_{\rm all}\leq I_1.
\label{eq:qI1}
\end{equation}
So, $p_{\rm all}$ uses a more demanding condition than $I_1$. The index $I_1$ uses the average number of members consulted. The quantity $p_{\rm all}$ counts only the decisions in which every member was consulted.

The all-member probability is directly related to the threshold index. At the 100 per cent threshold, $q_{100}=m$. Hence, from the definition of $I_3(x)$,
\begin{equation}
\boxed{I_3(100)=P(K\geq m)=P(K=m)=p_{\rm all}.}
\label{eq:I3_100_pall}
\end{equation}
Thus, $I_3(100)$ is both a threshold index value and the observed probability that a randomly selected decision involves all members. This connects the consultation index directly to the probability model used below. The index $I_1$ gives the average consultation breadth, while $I_3(100)$ gives the probability of reaching the maximum consultation level. Since $I_2=I_1$, neither $I_1$ nor $I_2$ alone determines the all-member probability.

The threshold function can also provide useful information when direct observations of all-member consultation are unavailable or too few. Since $I_3(x)$ is non-increasing with $x$, for any $x<100$,
\begin{equation}
I_3(100)\leq I_3(x).
\label{eq:pall_upper_bound}
\end{equation}
So, because $p_{\rm all}=I_3(100)$,
\begin{equation}
\boxed{p_{\rm all}\leq I_3(x),\qquad 0<x<100.}
\label{eq:pall_upper_bound_I3}
\end{equation}
For example, a recorded value of $I_3(90\%)$ gives an immediate upper bound on the probability of complete participation. In the two examples considered here, $I_3(90\%)=9/225=0.0400$ for the small group and $I_3(90\%)=37/1062\approx0.03484$ for the large group. Thus, even without using the direct all-member counts, the 90\% threshold alone would give the bounds $p_{\rm all}\leq0.0400$ and $p_{\rm all}\leq0.03484$, respectively. Estimates of $I_3(80\%)$ and $I_3(70\%)$ provide progressively weaker upper bounds. Such estimates may also come from informed assessments or group members' perceptions when detailed decision-wise records are unavailable. Their interpretation then depends on the reliability of those assessments. These higher-threshold values do not, by themselves, give a unique value of $p_{\rm all}$. An additional assumption about the upper tail between the highest available threshold and $100\%$ would be needed to estimate $I_3(100)$ from them.

This is useful when all-member consultation is rare. Even when direct measurement is difficult, a high-threshold value can give an upper bound on $p_{\rm all}$. If only an upper bound $u$ on $p_{\rm all}$ is available, then, under the same assumptions, the expected number $Y$ in $N'$ future decisions satisfies
\begin{equation}
E[Y]=N'p_{\rm all}\leq N'u.
\end{equation}
Moreover, for any integer $y\geq1$, monotonicity of the binomial upper tail in $p_{\rm all}$ gives
\begin{equation}
P(Y\geq y)\leq
\sum_{j=y}^{N'}\binom{N'}{j}u^j(1-u)^{N'-j}.
\label{eq:binomial_upper_bound_from_pall}
\end{equation}
So, a high-threshold estimate can provide both an upper bound on the all-member consultation rate and a corresponding conditional upper bound on future all-member events.

For the small group,
\begin{equation}
p_{\rm all}=\frac{n_{22}}{225}=\frac{1}{225}\approx0.00444.
\end{equation}
So, 0.444\% of the observed decisions involved all-member consultation. For the large group,
\begin{equation}
p_{\rm all}=\frac{n_{105}}{1062}=\frac{2}{1062}\approx0.00188.
\end{equation}
So, 0.188\% of the observed decisions involved all-member consultation.

\subsection{Binomial probability model}

We can use the observed value of $p_{\rm all}$ to examine possible outcomes for future decisions, assuming that the same conditions continue. The calculation uses three assumptions: the same coordinator and group continue, the probability of all-member consultation remains equal to the observed $p_{\rm all}$, and future decisions are independent. Under these assumptions, the events form independent Bernoulli trials with the same probability in each trial. These are simplifying assumptions, and they need not hold in an actual group.

Let $N^{\prime}$ be the number of future decisions, and let $Y$ be the number of these decisions in which all $m$ members are consulted. For this calculation, each future decision has two possible outcomes: all $m$ members are consulted, or fewer than $m$ members are consulted. Under the assumptions stated above, each decision is a Bernoulli trial. The number $Y$ of decisions with all-member consultation in $N^{\prime}$ trials therefore follows a binomial distribution. The probability that exactly $y$ of the $N^{\prime}$ decisions involve all-member consultation is
\begin{equation}
P(Y=y)=\binom{N^{\prime}}{y}p_{\rm all}^{\,y}(1-p_{\rm all})^{N^{\prime}-y}.
\label{eq:binomial}
\end{equation}
Here:
\begin{itemize}
\item $N^{\prime}$ is the number of future decisions being considered;
\item $y$ is the number of those decisions in which all $m$ members are consulted, where $0\leq y\leq N^{\prime}$;
\item $p_{\rm all}$ is the probability of an all-member consultation decision;
\item $1-p_{\rm all}$ is the probability that fewer than $m$ members are consulted;
\item $\binom{N^{\prime}}{y}=N^{\prime}!/[y!(N^{\prime}-y)!]$ is the number of different ways in which the $y$ decisions with all-member consultation can occur among the $N^{\prime}$ decisions;
\item $p_{\rm all}^{\,y}$ is the probability factor for the $y$ decisions with all-member consultation; and
\item $(1-p_{\rm all})^{N^{\prime}-y}$ is the probability factor for the remaining decisions, in which fewer than $m$ members are consulted.
\end{itemize}
The formula gives the probability of exactly $y$ all-member consultation decisions. The fraction $y/N^{\prime}$ gives the corresponding proportion of decisions, and $100y/N^{\prime}$ gives the percentage. Thus, when $N^{\prime}=100$, the value of $y$ is numerically equal to the percentage of decisions with all-member consultation.

For a simple numerical illustration, take $N^{\prime}=100$. The probabilities for $y=0,1,2,3,4$ and $5$ are given in Table~\ref{tab:binomial}. The table gives the probability of \emph{exactly} each value. These same values of $y$ are used in the Poisson comparison later in this section, so that the two probability models can be compared directly.

\begin{table}[htbp]
\centering
\caption{Binomial probability that exactly $y$ out of $N^{\prime}=100$ future decisions involve all members, using the observed value of $p_{\rm all}=n_m/N$ for each surveyed group.}
\label{tab:binomial}
\small
\begin{tabular}{c c c}
\toprule
$y$ & Small group & Large group\\
\midrule
0 & 0.640546 & 0.828199\\
1 & 0.285958 & 0.156264\\
2 & 0.063192 & 0.014594\\
3 & 0.009215 & 0.000900\\
4 & 0.000998 & $4.12\times10^{-5}$\\
5 & $8.55\times10^{-5}$ & $1.49\times10^{-6}$\\
\bottomrule
\end{tabular}
\vspace{1mm}
\begin{minipage}{0.94\textwidth}
\footnotesize
\textit{Note:} For this calculation, each future decision-making event is treated as an independent Bernoulli trial. The probability of consulting all members, $p_{\rm all}$, is assumed to be the same from one decision to the next. This probability is estimated from the group's documented information about decisions during the preceding five-year period. Thus, the table gives a conditional forecast based on the observed five-year rate and the stated independence and unchanged-probability assumptions; it is not an unconditional prediction of future behaviour.
\end{minipage}
\end{table}

The same probability interpretation applies to any fixed consultation threshold, not only the 100\% threshold. If a future decision is counted when at least $x\%$ of the members are consulted, the observed success probability is $I_3(x)$. If this probability remains unchanged and future decisions are independent, the number $Y_x$ of threshold-reaching decisions in $N^{\prime}$ future decisions follows
\begin{equation}
Y_x\sim\operatorname{Binomial}\!\left(N^{\prime},I_3(x)\right).
\label{eq:binomial_threshold}
\end{equation}
For the special case $x=100\%$, this reduces to the all-member model above because $I_3(100)=p_{\rm all}$. Thus, the binomial calculation can be viewed as a direct probability application of the threshold index at the maximum consultation level.

\subsection{Expected count and standard deviation}

For values of $y$ much larger than the expected count, the binomial probabilities become extremely small because the observed value of $p_{\rm all}$ is small. For a binomial random variable, the mean and variance are
\begin{equation}
E[Y]=N^{\prime}p_{\rm all},\qquad
\operatorname{Var}(Y)=N^{\prime}p_{\rm all}(1-p_{\rm all}),
\end{equation}
so that the standard deviation is
\begin{equation}
\sigma_Y=\sqrt{N^{\prime}p_{\rm all}(1-p_{\rm all})}.
\end{equation}
These formulas give the expected number of all-member consultation decisions and the corresponding standard deviation when the probability stays unchanged.

Table~\ref{tab:binomial_moments} gives the expectation and standard deviation for three future decision horizons. The shortest horizon is the same as in Table~\ref{tab:binomial}. The two longer horizons show how the expected number and the relative spread change when the same observed rate is applied to more future decisions.

\begin{table}[htbp]
\centering
\caption{Expectation and standard deviation of the number of decisions with all-member consultation under the binomial model. The observed value of $p_{\rm all}$ is kept unchanged.}
\label{tab:binomial_moments}
\small
\begin{tabular}{c c c c c}
\toprule
Future decisions & \multicolumn{2}{c}{Small group} & \multicolumn{2}{c}{Large group}\\
\cmidrule(lr){2-3}\cmidrule(lr){4-5}
$N^{\prime}$ & $E[Y]$ & $\sigma_Y$ & $E[Y]$ & $\sigma_Y$\\
\midrule
100 & 0.4444 & 0.6652 & 0.1883 & 0.4336\\
10,000 & 44.4444 & 6.6518 & 18.8324 & 4.3355\\
1,000,000 & 4444.4444 & 66.5184 & 1883.2392 & 43.3554\\
\bottomrule
\end{tabular}
\end{table}

\subsection{Poisson illustration for a future five-year period}

Under the stated assumptions, the binomial model gives the exact probability. When the all-member consultation probability is small and the number of future decisions is large, we can approximate the binomial distribution by a Poisson distribution. This approximation is useful here because the observed all-member consultation probabilities are small.

Let $Y$ again denote the number of all-member consultation decisions in a future set of $N^{\prime}$ decisions. The Poisson approximation uses
\begin{equation}
Y\sim\operatorname{Poisson}(\lambda),\qquad
\lambda=N^{\prime}p_{\rm all},
\label{eq:poisson}
\end{equation}
where $\lambda$ is the expected number of all-member consultation decisions. Since $p_{\rm all}=I_3(100)$, the Poisson mean can also be written as
\begin{equation}
\boxed{\lambda=N^{\prime}I_3(100).}
\label{eq:poisson_I3}
\end{equation}
Thus, the threshold-index value that measures the observed frequency of complete consultation also gives the mean of the future Poisson count, provided that the probability is assumed to remain unchanged. The probability of exactly $y$ such decisions is
\begin{equation}
P(Y=y)=\frac{e^{-\lambda}\lambda^y}{y!},
\qquad y=0,1,2,\ldots.
\label{eq:poisson_probability}
\end{equation}
For a specified minimum number $y_0$, the probability of at least $y_0$ all-member decisions is
\begin{equation}
P(Y\geq y_0)=1-\sum_{y=0}^{y_0-1}\frac{e^{-\lambda}\lambda^y}{y!}.
\label{eq:poisson_tail}
\end{equation}
More generally, for a fixed threshold $x$ for which $I_3(x)$ is small, the corresponding count can be approximated by
\begin{equation}
Y_x\approx\operatorname{Poisson}\!\left(N^{\prime}I_3(x)\right).
\label{eq:poisson_threshold}
\end{equation}
The all-member case used in this section is the special case $x=100\%$.

We can compare the Poisson calculation with the binomial calculation using the same number of future decisions. For $N^{\prime}=100$, the Poisson means are $\lambda=0.4444$ for the small group and $\lambda=0.1883$ for the large group. The two sets of probabilities are close in these examples because $p_{\rm all}$ is small. The difference becomes more noticeable in relative terms for the smaller tail probabilities, even though the absolute probabilities remain very small. Table~\ref{tab:binomial_poisson_compare} gives the exact binomial and Poisson probabilities for $y=0,1,2,3,4$ and $5$. The comparison is made for the same 100 future decisions used in Table~\ref{tab:binomial}.

\begin{table}[htbp]
\centering
\caption{Comparison of exact binomial and Poisson probabilities for the same $N^{\prime}=100$ future decisions. The Poisson mean is $\lambda=N^{\prime}p_{\rm all}$.}
\label{tab:binomial_poisson_compare}
\small
\begin{tabular}{c cc cc}
\toprule
& \multicolumn{2}{c}{Small group} & \multicolumn{2}{c}{Large group}\\
\cmidrule(lr){2-3}\cmidrule(lr){4-5}
$y$ & Binomial & Poisson & Binomial & Poisson\\
\midrule
0 & 0.640546 & 0.641180 & 0.828199 & 0.828346\\
1 & 0.285958 & 0.284969 & 0.156264 & 0.155997\\
2 & 0.063192 & 0.063326 & 0.014594 & 0.014689\\
3 & 0.009215 & 0.009382 & 0.000900 & 0.000922\\
4 & 0.000998 & 0.001042 & $4.12\times10^{-5}$ & $4.34\times10^{-5}$\\
5 & $8.55\times10^{-5}$ & $9.27\times10^{-5}$ & $1.49\times10^{-6}$ & $1.64\times10^{-6}$\\
\bottomrule
\end{tabular}
\end{table}

We now use the actual number of decisions in the preceding five-year period for each of the two detailed groups. This gives a simple five-year illustration. We ask what the Poisson model would give if the group made the same number of decisions in a future five-year period and if $p_{\rm all}$ remained unchanged. This is a conditional forecast based on the observed five-year rate and the stated assumptions, not an unconditional prediction of the group's future behaviour. For the small group, $N_5=225$ and $p_{\rm all}=1/225$, giving
\begin{equation}
\lambda_5=N_5p_{\rm all}=225\times\frac{1}{225}=1.
\end{equation}
For the large group, $N_5=1062$ and $p_{\rm all}=2/1062$, giving
\begin{equation}
\lambda_5=N_5p_{\rm all}=1062\times\frac{2}{1062}=2.
\end{equation}
So, under these assumptions, the expected numbers of all-member consultation decisions in the next five years are 1 and 2 for the small and large groups, respectively.

To examine different levels of future all-member consultation, let $z$ be a percentage of the total number of decisions in the next five years. The smallest whole number of all-member decisions corresponding to at least $z\%$ is
\begin{equation}
y_z=\left\lceil\frac{z}{100}N_5\right\rceil.
\label{eq:poisson_threshold_count}
\end{equation}
We consider $z=0.5\%,1\%,2\%,5\%$ and $10\%$. Table~\ref{tab:poisson_fiveyear} gives the probability that at least the corresponding number of decisions will involve all members. We calculate the values from Eq.~\eqref{eq:poisson_tail}.

\begin{table}[htbp]
\centering
\caption{Poisson probability that at least the stated percentage of decisions in the next five years involve all members, assuming the same number of decisions as in the preceding five-year period and the same observed $p_{\rm all}$.}
\label{tab:poisson_fiveyear}
\small
\begin{tabular}{c cc cc}
\toprule
& \multicolumn{2}{c}{Small group ($N_5=225$, $\lambda_5=1$)} & \multicolumn{2}{c}{Large group ($N_5=1062$, $\lambda_5=2$)}\\
\cmidrule(lr){2-3}\cmidrule(lr){4-5}
$z$ & $y_z$ & $P(Y\geq y_z)$ & $y_z$ & $P(Y\geq y_z)$\\
\midrule
0.5\% & 2 & 0.26424 & 6 & 0.016564\\
1\% & 3 & 0.080301 & 11 & $8.31\times10^{-6}$\\
2\% & 5 & 0.003660 & 22 & $5.53\times10^{-16}$\\
5\% & 12 & $8.32\times10^{-10}$ & 54 & $1.10\times10^{-56}$\\
10\% & 23 & $1.48\times10^{-23}$ & 107 & $1.82\times10^{-141}$\\
\bottomrule
\end{tabular}
\vspace{1mm}
\begin{minipage}{0.94\textwidth}
\footnotesize
\textit{Note:} The table is a conditional calculation. It assumes that the observed $p_{\rm all}$ remains unchanged, that future decisions are independent, and that the number of decisions in the next five years is the same as in the preceding five-year period for the corresponding group. The percentages refer to all-member consultation decisions, not to the consultation index $I_1$.
\end{minipage}
\end{table}

These probabilities illustrate the implications of the observed all-member consultation rates over a five-year period. They are not unconditional predictions of future behaviour. If the number of decisions or the all-member consultation rate changes, the Poisson mean will also change. The calculation is therefore conditional on the same assumptions as the binomial model.

\subsection{Dependence on the number of future decisions}

For 100 future decisions, the expected number is about 0.44 for the small group and about 0.19 for the large group. Therefore, values such as 30, 40 or 50 decisions with all-member consultation in 100 decisions are far above the expected number under this assumption. Their exact probabilities are correspondingly extremely small. For 10,000 future decisions, the expected numbers are about 44.4 and 18.8, respectively. For 1,000,000 future decisions, the expected numbers are about 4444.4 and 1883.2. These values do not mean that such numbers will occur. They are the averages of the corresponding binomial distributions under the stated assumptions. The standard deviations in Table~\ref{tab:binomial_moments} show the typical spread around these averages. The mean increases in direct proportion to the number of future decisions, while the standard deviation increases only as its square root. To see the relative change, divide the standard deviation by the mean. From the expressions above,
\begin{equation}
\frac{\sigma_Y}{E[Y]}=\sqrt{\frac{1-p_{\rm all}}{N^{\prime}p_{\rm all}}}.
\label{eq:relative_binomial_spread}
\end{equation}
So, for a fixed $p_{\rm all}$, the relative standard deviation decreases as $1/\sqrt{N^{\prime}}$. For example, when $N^{\prime}$ increases from 10,000 to 1,000,000, the expected count increases by a factor of 100, while the standard deviation increases by a factor of 10. The relative variation therefore becomes smaller, even though the absolute standard deviation becomes larger.

The results in Table~\ref{tab:binomial} are theoretical conditional probabilities. They assume that $p_{\rm all}$ remains unchanged and that successive decisions are independent. They are therefore conditional forecasts rather than unconditional predictions of future behaviour. The observed data are aggregate data for a preceding five-year period. They do not give a year-by-year record of changes in the consultation rate. The binomial calculation therefore does not model a time trend. It asks what would happen in a future set of decisions if the observed probability remained unchanged. The future probability may change if the coordinator, the members, the issues, the decision process or the pattern of consultation changes. An unconditional forecast would require data recorded over time and a model that allows the consultation probability to change.

The calculation is still useful because it shows how the observed all-member consultation rate limits the chance of future all-member decisions when the rate remains unchanged. It also gives a simple comparison between the two groups.

\section{Member-wise participation and network representation}
\subsection{Illustrative member assignment for the small group}

Table~\ref{tab:table1} gives only $n_k$; it does not say which particular members were consulted. Therefore, the individual values of $F_i$ cannot be determined from Table~\ref{tab:table1} without an additional assumption.

The member-wise values below are illustrative, not observed. We construct them from the aggregate data only to demonstrate the calculation of the Gini coefficient, Lorenz curve and Shannon index. They do not describe the actual consultation histories of particular members.

For illustration, suppose that whenever a decision involves $k$ members, the consulted members are members $1,2,\ldots,k$. This gives a convenient nested assignment from which member-wise values can be calculated. This is only a construction based on the aggregate counts; it does not imply that these particular members were selected in this way in the surveyed group. With this assignment,
\begin{equation}
F_i=\sum_{k=i}^{m}n_k.
\label{eq:Fi}
\end{equation}
Table~\ref{tab:table3} gives the resulting values.

\begin{table}[htbp]
\centering
\caption{Member-wise participation frequencies from the illustrative nested assignment in Eq.~\eqref{eq:Fi}, constructed from the aggregate data in Table~\ref{tab:table1}. The assignment is illustrative and is not an observed individual consultation history.}
\label{tab:table3}
\small
\begin{tabular}{r r r}
\toprule
$i$ & $F_i$ & $f_i=F_i/N$\\
\midrule
1 & 225 & 1.0000\\
2 & 222 & 0.9867\\
3 & 217 & 0.9644\\
4 & 209 & 0.9289\\
5 & 199 & 0.8844\\
6 & 186 & 0.8267\\
7 & 171 & 0.7600\\
8 & 153 & 0.6800\\
9 & 134 & 0.5956\\
10 & 116 & 0.5156\\
11 & 100 & 0.4444\\
12 & 85 & 0.3778\\
13 & 71 & 0.3156\\
14 & 58 & 0.2578\\
15 & 47 & 0.2089\\
16 & 37 & 0.1644\\
17 & 28 & 0.1244\\
18 & 21 & 0.0933\\
19 & 14 & 0.0622\\
20 & 9 & 0.0400\\
21 & 4 & 0.0178\\
22 & 1 & 0.0044\\
\bottomrule
\end{tabular}
\end{table}

Using Eq.~\eqref{eq:I2}, these values give
\begin{equation*}
I_2=\frac{1}{22}\sum_{i=1}^{22}f_i=\frac{2307}{22\times225}\approx0.4661,
\end{equation*}
which agrees with the value calculated from the decision-based data. The particular values of $F_i$ depend on the assignment used to construct the member-wise example. Their sum is fixed by Eq.~\eqref{eq:counting}.

\subsection{Illustrative member assignment for the large group}

We apply the same illustrative nested assignment to the second survey-based aggregate. If the consulted members for a decision involving $k$ members are members $1,2,\ldots,k$, Eq.~\eqref{eq:Fi} gives the corresponding member-wise frequencies. Table~\ref{tab:table4} shows these values. As in the small-group case, this is an illustrative construction; the aggregate data in Table~\ref{tab:table2} do not determine the individual values uniquely.

\begin{table}[htbp]
\centering
\caption{Member-wise participation frequencies from the illustrative nested assignment in Eq.~\eqref{eq:Fi}, constructed from the aggregate data in Table~\ref{tab:table2}. The assignment is illustrative and is not an observed individual consultation history.}
\label{tab:table4}
\scriptsize
\begin{tabular}{r r r @{\qquad} r r r}
\toprule
$i$ & $F_i$ & $f_i=F_i/N$ & $i$ & $F_i$ & $f_i=F_i/N$\\
\midrule
1 & 1060 & 0.9981 & 54 & 438 & 0.4124\\
2 & 1058 & 0.9962 & 55 & 423 & 0.3983\\
3 & 1055 & 0.9934 & 56 & 409 & 0.3851\\
4 & 1052 & 0.9906 & 57 & 395 & 0.3719\\
5 & 1049 & 0.9878 & 58 & 381 & 0.3588\\
6 & 1045 & 0.9840 & 59 & 367 & 0.3456\\
7 & 1041 & 0.9802 & 60 & 354 & 0.3333\\
8 & 1037 & 0.9765 & 61 & 341 & 0.3211\\
9 & 1032 & 0.9718 & 62 & 328 & 0.3089\\
10 & 1027 & 0.9670 & 63 & 315 & 0.2966\\
11 & 1021 & 0.9614 & 64 & 303 & 0.2853\\
12 & 1014 & 0.9548 & 65 & 291 & 0.2740\\
13 & 1007 & 0.9482 & 66 & 279 & 0.2627\\
14 & 999 & 0.9407 & 67 & 267 & 0.2514\\
15 & 990 & 0.9322 & 68 & 256 & 0.2411\\
16 & 981 & 0.9237 & 69 & 245 & 0.2307\\
17 & 972 & 0.9153 & 70 & 234 & 0.2203\\
18 & 962 & 0.9058 & 71 & 223 & 0.2100\\
19 & 952 & 0.8964 & 72 & 212 & 0.1996\\
20 & 942 & 0.8870 & 73 & 202 & 0.1902\\
21 & 931 & 0.8766 & 74 & 192 & 0.1808\\
22 & 920 & 0.8663 & 75 & 182 & 0.1714\\
23 & 909 & 0.8559 & 76 & 172 & 0.1620\\
24 & 897 & 0.8446 & 77 & 163 & 0.1535\\
25 & 885 & 0.8333 & 78 & 154 & 0.1450\\
26 & 873 & 0.8220 & 79 & 145 & 0.1365\\
27 & 860 & 0.8098 & 80 & 136 & 0.1281\\
28 & 847 & 0.7976 & 81 & 128 & 0.1205\\
29 & 834 & 0.7853 & 82 & 120 & 0.1130\\
30 & 820 & 0.7721 & 83 & 112 & 0.1055\\
31 & 806 & 0.7589 & 84 & 104 & 0.0979\\
32 & 792 & 0.7458 & 85 & 97 & 0.0913\\
33 & 778 & 0.7326 & 86 & 90 & 0.0847\\
34 & 763 & 0.7185 & 87 & 83 & 0.0782\\
35 & 748 & 0.7043 & 88 & 76 & 0.0716\\
36 & 733 & 0.6902 & 89 & 70 & 0.0659\\
37 & 717 & 0.6751 & 90 & 64 & 0.0603\\
38 & 701 & 0.6601 & 91 & 58 & 0.0546\\
39 & 684 & 0.6441 & 92 & 52 & 0.0490\\
40 & 667 & 0.6281 & 93 & 47 & 0.0443\\
41 & 650 & 0.6121 & 94 & 42 & 0.0395\\
42 & 633 & 0.5960 & 95 & 37 & 0.0348\\
43 & 615 & 0.5791 & 96 & 32 & 0.0301\\
44 & 598 & 0.5631 & 97 & 28 & 0.0264\\
45 & 581 & 0.5471 & 98 & 24 & 0.0226\\
46 & 564 & 0.5311 & 99 & 20 & 0.0188\\
47 & 547 & 0.5151 & 100 & 16 & 0.0151\\
48 & 531 & 0.5000 & 101 & 13 & 0.0122\\
49 & 515 & 0.4849 & 102 & 10 & 0.0094\\
50 & 499 & 0.4699 & 103 & 7 & 0.0066\\
51 & 483 & 0.4548 & 104 & 4 & 0.0038\\
52 & 468 & 0.4407 & 105 & 2 & 0.0019\\
53 & 453 & 0.4266 & & & \\
\bottomrule
\end{tabular}
\end{table}

Using Eq.~\eqref{eq:I2}, the values in Table~\ref{tab:table4} give
\begin{equation*}
I_2=\frac{1}{105}\sum_{i=1}^{105}f_i=\frac{52341}{105\times1062}\approx0.4694,
\end{equation*}
which is the same as the value obtained from the decision-based calculation for Table~\ref{tab:table2}.

For the two survey-based aggregates, Tables~\ref{tab:table3} and~\ref{tab:table4} show the member-wise frequencies under the stated nested assignment. In real data, the values of $F_i$ would come from records that identify which members were consulted for each decision. We use the nested assignments only because the aggregate tables give the number of decisions involving $k$ members, not the identities of the members consulted.

\subsection{A useful bound on member-wise inequality}

The nested assignment used above has a useful mathematical interpretation. It is not only one possible assignment of the consulted members; it gives the most unequal member-wise distribution that can be obtained from the fixed decision-side counts.

To see this, arrange the decision counts in any order and define
\begin{equation}
h_j=\#\{r:k_r\geq j\},
\qquad j=1,2,\ldots,m.
\label{eq:hj}
\end{equation}
Thus, $h_j$ is the number of decisions in which at least $j$ members are consulted. Under the nested assignment, Eq.~\eqref{eq:Fi} gives
\begin{equation}
F_j=h_j.
\label{eq:nested_h}
\end{equation}
The vector $(h_1,\ldots,h_m)$ is therefore the member-wise vector produced by the nested construction.

More generally, suppose that the member-wise counts are arranged in decreasing order,
\[
F_{(1)}\geq F_{(2)}\geq\cdots\geq F_{(m)}.
\]
A member-wise vector is possible for the given decision-side sequence $\{k_r\}$ if and only if its components satisfy,
\begin{equation}
\sum_{i=1}^{m}F_{(i)}
=
\sum_{r=1}^{N}k_r
\end{equation}
and
\begin{equation}
\sum_{i=1}^{t}F_{(i)}
\leq
\sum_{j=1}^{t}h_j
=
\sum_{r=1}^{N}\min(k_r,t),
\qquad t=1,\ldots,m.
\label{eq:majorization_bound}
\end{equation}
These are the standard Gale--Ryser/Ryser-type conditions for a zero--one matrix with prescribed row sums and column sums \cite{ryser1957}. In the present setting, the rows are decisions and the columns are members. Thus, Eq.~\eqref{eq:majorization_bound} describes how the member-wise counts can vary when the decision-side counts are fixed. In simple terms, the consultation events cannot be concentrated among a small set of members beyond what the decision-side counts allow.

The nested vector $(h_1,\ldots,h_m)$ is the majorization-maximal member-wise vector, apart from a permutation of the member labels. It therefore gives the largest possible inequality among members. Hence,
\begin{equation}
G_{\max}=G(h_1,\ldots,h_m),
\qquad
S_{F,\min}=S_F(h_1,\ldots,h_m).
\label{eq:extreme_nested}
\end{equation}
At the other end, let $L$ be the total number of consultation events. Then
\begin{equation}
L=\sum_{r=1}^{N}k_r=q_{\mathrm{bal}}m+r_0,
\qquad 0\leq r_0<m.
\end{equation}
The most even integer distribution has $r_0$ members with $q_{\mathrm{bal}}+1$ consultation events and the remaining $m-r_0$ members with $q_{\mathrm{bal}}$ events. This distribution is attainable for the same decision-side counts. It therefore gives the minimum possible Gini coefficient and the maximum possible Shannon participation index. In particular,
\begin{equation}
G_{\min}=\frac{r_0(m-r_0)}{mL}.
\label{eq:gini_min_fixed}
\end{equation}
The maximum value of $S_F$ is obtained by using these balanced counts in Eq.~\eqref{eq:SF}. So, for a fixed set of decision-side counts, both measures have definite attainable ranges:
\begin{equation}
G_{\min}\leq G\leq G_{\max},
\qquad
S_{F,\min}\leq S_F\leq S_{F,\max}.
\label{eq:inequality_ranges}
\end{equation}

For the small-group aggregate, these bounds are
\[
0.0011\leq G\leq0.4303,
\qquad
0.8933\leq S_F\leq0.999998.
\]
For the large-group aggregate, they are
\[
0.0005\leq G\leq0.4200,
\qquad
0.9325\leq S_F\leq0.9999999.
\]
The upper Gini values and lower Shannon values are exactly those obtained from the nested assignments in Tables~\ref{tab:table3} and~\ref{tab:table4}. Thus, the nested examples used in this paper represent the most unequal member-wise realisations compatible with the two aggregate decision-side datasets. The simulations can be viewed as examples of other possible allocations within the same overall consultation level. We use this result only to identify the attainable range and the two extreme cases; a full combinatorial treatment of all possible assignments is beyond the scope of this paper.

\subsection{Bipartite network representation}

The matrix $x_{ir}$ gives another simple view of the same records. It has one row for each member and one column for each decision. A 1 means that the member is consulted, and a 0 means that the member is not consulted. The same information can be drawn as a network. Each member is one node, and each decision is another node. A link joins member $i$ and decision $r$ when $x_{ir}=1$.

This is a \emph{bipartite network}. In physics and network science, its nodes are divided into two groups, with links allowed only between the two groups \cite{newman2010}. Here, the two groups are members and decisions. There are no links between two members or between two decisions. A link exists only when a member is consulted for a decision.

For example, consider three members, $u$, $v$ and $w$, and two decisions, 1 and 2. Suppose $u$ and $v$ are consulted for decision 1, while $v$ and $w$ are consulted for decision 2. The corresponding matrix is
\begin{equation*}
\begin{array}{c|cc}
 & \text{Decision 1} & \text{Decision 2}\\
\hline
u & 1 & 0\\
v & 1 & 1\\
w & 0 & 1
\end{array}
\end{equation*}
There are four member--decision links out of a possible six. The fraction of possible links that are present is $4/6$. From Eq.~\eqref{eq:I2} this is also
\begin{equation*}
I_2=\frac{1+0+1+1+0+1}{3\times2}=\frac{4}{6}.
\end{equation*}
This example shows the network description directly, without requiring any additional network theory. In general, there are $mN$ possible member--decision links, and Eq.~\eqref{eq:I1} gives the fraction of these links that are actually present. Thus, $I_1$ can also be viewed as the density (or filling fraction) of the member--decision bipartite network.

This view also shows a limitation of $I_1$. The value of $I_1$ gives the total number of member--decision links relative to the number possible, but it does not show how these links are distributed among members. Two groups can have the same number of consultation events. In one group the events may be shared almost equally. In another, most events may involve only a few members. The two groups can then have the same $I_1$, even though their participation distributions differ.

\subsubsection{Application of the network representation}

When the identities of the consulted members are available, we can form the matrix $x_{ir}$ directly from the records. The row sums give the member-side frequencies. The column sums give the decision-side consultation counts:
\begin{equation}
F_i=\sum_{r=1}^{N}x_{ir},
\qquad
k_r=\sum_{i=1}^{m}x_{ir}.
\label{eq:network_degrees}
\end{equation}
The total number of member--decision links is therefore
\begin{equation}
L=\sum_{i=1}^{m}F_i=\sum_{r=1}^{N}k_r.
\label{eq:network_links}
\end{equation}
Since there are $mN$ possible links, the network density is
\begin{equation}
\frac{L}{mN}=I_1=I_2.
\label{eq:network_density}
\end{equation}
So, the network representation does not give a new overall index. It is another way to represent the same consultation events used in the definitions of $I_1$ and $I_2$. The row sums give the Gini coefficient and Shannon participation index, while the column sums give $I_3(x)$. A complete member--decision record can therefore provide all these quantities when member identities are available.

The aggregate $n_k$ data give only the distribution of the column sums, so they do not determine the row sums uniquely. Therefore, the network representation and the member-wise measures require more detailed records than the aggregate data used in Tables~\ref{tab:table1} and~\ref{tab:table2}.

\subsection{Participation inequality among members: Gini coefficient and Shannon index}

The member-wise quantities $F_i$ provide information that $I_1$ does not. The index $I_1$ measures the overall breadth of consultation across decisions, whereas $F_i$ records how often member $i$ is consulted. Two groups can therefore have the same $I_1$ but different member-wise distributions. The Gini coefficient, Lorenz curve and Shannon participation index describe this difference.

A standard measure of this inequality is the Gini coefficient. The Gini coefficient and Lorenz curve are often used together to describe inequality in a distribution \cite{lorenz1905,gini2012}. We apply the standard definition to the member-wise consultation counts to measure how unevenly consultation is distributed. For the $m$ values $F_1,F_2,\ldots,F_m$, define
\begin{equation}
G=\frac{1}{2m^2\bar F}\sum_{i=1}^{m}\sum_{j=1}^{m}|F_i-F_j|,
\label{eq:gini}
\end{equation}
where
\begin{equation}
\bar F=\frac{1}{m}\sum_{i=1}^{m}F_i=NI_1.
\label{eq:Fmean}
\end{equation}
When total participation is nonzero, the Gini coefficient satisfies $0\leq G\leq1$. It is zero when all members have the same consultation count. It becomes larger as the consultation counts become more unequal. If no member is consulted in any decision, then $I_1=0$ and all $F_i=0$. In that case, the denominator in Eq.~\eqref{eq:gini} is zero, so the Gini coefficient is not defined.

The Gini coefficient does not measure the overall amount of consultation. It measures inequality in the consultation counts across members. A higher value of $G$ indicates greater concentration among a few members, while a lower value indicates a more even distribution. Two groups can therefore have the same overall breadth (represented by $I_1$) but different Gini values. A statement such as twenty percent of members account for eighty percent of consultation events describes concentration, but it is not itself the Gini coefficient. The Gini coefficient uses the full set of member-wise consultation counts.

For the illustrative member assignments in Tables~\ref{tab:table3} and~\ref{tab:table4}, the Gini coefficients are about
\begin{equation*}
G\approx0.4303\qquad\text{and}\qquad G\approx0.4200,
\end{equation*}
respectively. As shown above, these are the maximum attainable Gini coefficients for the two fixed aggregate datasets. Other assignments can give smaller Gini coefficients while keeping the same $\sum_i F_i$ and hence the same $I_2=I_1$.

\begin{figure}[htbp]
\centering
\includegraphics[width=0.88\textwidth]{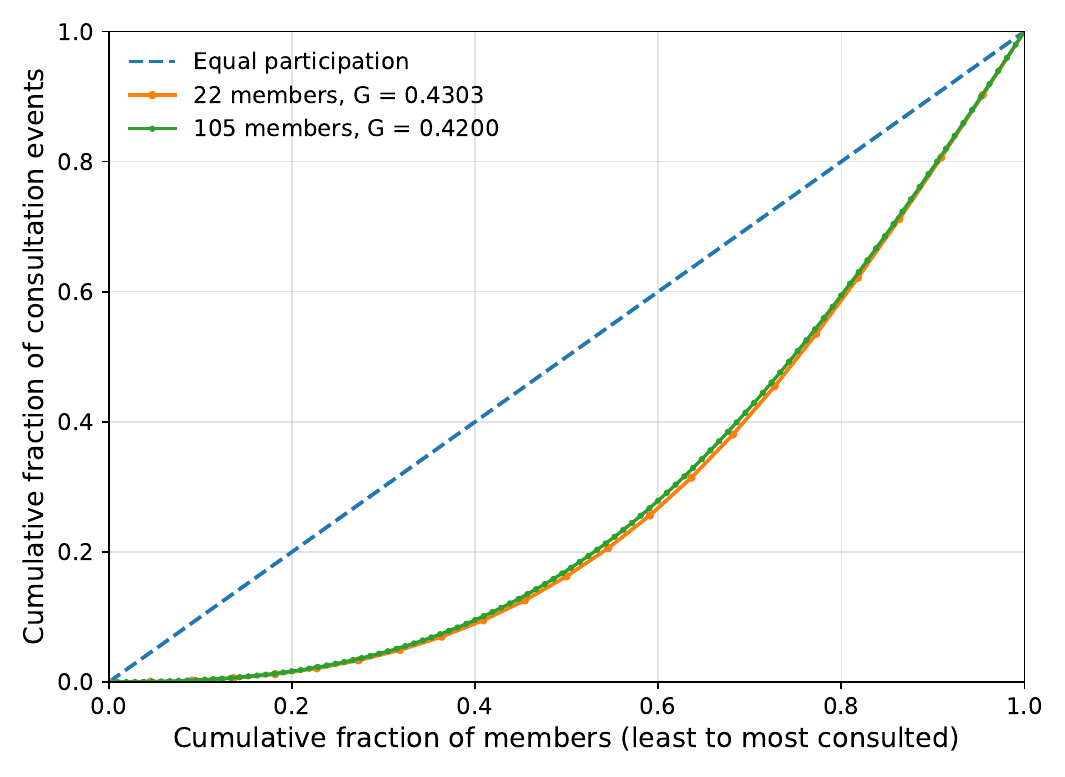}
\caption{Lorenz curves for the illustrative synthetic member assignments in Tables~\ref{tab:table3} and~\ref{tab:table4}. Members are ordered from the least consulted to the most consulted. The diagonal line represents equal participation. The assignments are constructed from aggregate consultation counts. The areas between the diagonal and the two curves give the corresponding Gini coefficients through Eq.~\eqref{eq:lorenzgini}.}
\label{fig:lorenz}
\end{figure}

The Gini coefficient can also be shown with a Lorenz curve. We first arrange the members from the least consulted to the most consulted. The horizontal axis gives the cumulative fraction of members. The vertical axis gives the cumulative fraction of consultation events received by them. Equal consultation gives the diagonal line $y=x$. A curve below this line shows unequal participation. The Gini coefficient is related to the area between the two curves. If $\mathcal{A}_{L}$ is the area under the Lorenz curve, then
\begin{equation}
G=1-2\mathcal{A}_{L}.
\label{eq:lorenzgini}
\end{equation}
For the illustrative data, we construct the Lorenz curves from the $F_i$ values in Tables~\ref{tab:table3} and~\ref{tab:table4} and show them in Fig.~\ref{fig:lorenz}. The two curves correspond to $G\approx0.4303$ and $G\approx0.4200$, respectively. The second curve lies closer to the equality line, consistent with its lower Gini coefficient and more even distribution of consultation among members.

These Lorenz curves are based on the illustrative member assignments, not directly on the aggregate $n_k$ values. The aggregate tables do not identify which members were consulted. Therefore, $F_i$, the Lorenz curve and the Gini coefficient cannot in general be obtained uniquely from $n_k$ alone.

For real data, records containing only $k$ for each decision are enough to calculate $I_1$, but they do not determine the Gini coefficient or Lorenz curve uniquely. If the records also identify which members were consulted, we can calculate $I_1$, the values of $F_i$, the Gini coefficient and the Lorenz curve. These measures answer different questions: $I_1$ gives the overall breadth of consultation, while $G$ and the Lorenz curve show how evenly consultation is distributed among members.

The same member-wise data can also be described with a Shannon index. It gives another measure of how evenly consultation events are shared among members. Unlike the Gini coefficient, for which a larger value means greater inequality, a larger Shannon participation index means greater evenness. If the total number of consultation events is nonzero, define
\begin{equation}
q_i=\frac{F_i}{\sum_{j=1}^{m}F_j},
\qquad i=1,2,\ldots,m,
\label{eq:qi}
\end{equation}
so that $q_i$ is the fraction of all consultation events received by member $i$. Terms with $q_i=0$ are taken to contribute zero. The Shannon entropy of this member-wise distribution is
\begin{equation}
H_F=-\sum_{i=1}^{m}q_i\ln q_i.
\label{eq:HF}
\end{equation}
We use the normalised form
\begin{equation}
S_F=\frac{H_F}{\ln m},
\qquad 0\leq S_F\leq1,
\qquad m>1.
\label{eq:SF}
\end{equation}
The value $S_F=1$ occurs when all members receive the same number of consultation events. Smaller values mean that the events are less evenly distributed, or more concentrated among fewer members. If all consultation events go to one member, $S_F=0$. When $I_1=0$, there are no consultation events. The distribution $q_i$ is then not defined. The Shannon participation index is also not defined in that case.

The Gini coefficient and the Shannon participation index use the same member-wise consultation counts but describe them in different ways. A larger Gini coefficient indicates greater inequality, while a larger Shannon participation index indicates greater evenness. Neither changes the overall breadth measured by $I_1$.

These three measures describe different aspects of the same consultation process. $I_1$ measures how widely members are consulted across decisions. The Gini coefficient measures how unequal the counts are across members. The Shannon participation index measures how evenly consultation is shared among members.

For the illustrative nested assignments used in Tables~\ref{tab:table3} and~\ref{tab:table4}, the normalised Shannon participation index is
\begin{equation*}
S_F\approx0.8933
\qquad\text{and}\qquad
S_F\approx0.9325,
\end{equation*}
respectively. As shown above, the Shannon values are the lowest that can be obtained from the given aggregate data. Other assignments can give larger values while keeping the same overall consultation level. The higher value in the second case is consistent with its lower Gini value and with its Lorenz curve being closer to the equality line.

\section{Indices from the fitted distribution}
The fitted distribution can also be used to calculate the indices. This lets us compare the values obtained from the aggregate data with those from the fitted model. We keep the $k=0$ part separate because we fit the maximum-entropy model only for $k\geq1$.

Let
\begin{equation}
Z(b,c)=\sum_{j=1}^{m}j^b e^{-cj}
\label{eq:Z}
\end{equation}
and let
\begin{equation}
\alpha=1-\frac{n_0}{N}.
\label{eq:alpha}
\end{equation}
Here $\alpha$ is the observed fraction of decisions with at least one member consulted. The fitted probability for $k\geq1$, when expressed relative to all decisions, is
\begin{equation}
P_{\rm fit}(k)=\alpha\frac{k^b e^{-ck}}{Z(b,c)},
\qquad k=1,2,\ldots,m.
\label{eq:Pfit}
\end{equation}
The remaining probability, $1-\alpha$, is assigned to $k=0$.

The mean number of consulted members obtained from the fitted distribution is
\begin{equation}
E_{\rm fit}=\alpha\frac{\displaystyle\sum_{k=1}^{m}k^{b+1}e^{-ck}}{Z(b,c)}.
\label{eq:Efit}
\end{equation}
Since $I_2=I_1$, the fitted member-side index is
\begin{equation}
\boxed{I_{2,\rm fit}=I_{1,\rm fit}
=\frac{\alpha}{m}
\frac{\displaystyle\sum_{k=1}^{m}k^{b+1}e^{-ck}}{Z(b,c)}}.
\label{eq:I2fit}
\end{equation}
So, the overall consultation index can be obtained from the fitted parameters and the observed zero-consultation fraction.

We can also obtain the threshold index from the fitted distribution. For
\begin{equation}
q_x=\left\lceil\frac{xm}{100}\right\rceil,
\end{equation}
we have
\begin{equation}
\boxed{I_{3,\rm fit}(x)=
\alpha\frac{\displaystyle\sum_{k=q_x}^{m}k^b e^{-ck}}{Z(b,c)}}.
\label{eq:I3fit}
\end{equation}
So, $I_3(x)$ is the upper-tail probability of the fitted consultation-level distribution.

The member-wise quantities need some care. The fitted distribution gives the number of decisions at each consultation level, but it does not say which members were consulted. To obtain a member-wise distribution from the fitted model, we therefore use the same nested assignment as in the illustrative member-wise calculations. With this assignment, the fitted expected number of consultation events received by member $i$ is
\begin{equation}
F_{i,\rm fit}=N\alpha
\frac{\displaystyle\sum_{k=i}^{m}k^b e^{-ck}}{Z(b,c)},
\qquad i=1,2,\ldots,m.
\label{eq:Fifit}
\end{equation}
For convenience, define the fitted member-wise tail
\begin{equation}
U_i(b,c)=
\frac{\displaystyle\sum_{k=i}^{m}k^b e^{-ck}}{Z(b,c)}.
\label{eq:Ui_fit}
\end{equation}
Then $F_{i,\rm fit}=N\alpha U_i(b,c)$. Hence
\begin{equation}
q_{i,\rm fit}=
\frac{U_i(b,c)}{\displaystyle\sum_{j=1}^{m}U_j(b,c)}.
\label{eq:qifit}
\end{equation}
Under the nested assignment, we can write the fitted Gini coefficient directly from the fitted distribution as
\begin{equation}
\boxed{
G_{\rm fit}=
\frac{\displaystyle\sum_{i=1}^{m}\sum_{j=1}^{m}
\left|U_i(b,c)-U_j(b,c)\right|}
{\displaystyle2m\sum_{i=1}^{m}U_i(b,c)}.
}
\label{eq:Gfit}
\end{equation}
The fitted normalised Shannon participation index is
\begin{equation}
\boxed{
S_{F,\rm fit}=-\frac{\displaystyle\sum_{i=1}^{m}q_{i,\rm fit}\ln q_{i,\rm fit}}{\ln m},
}
\label{eq:SFfit}
\end{equation}
where $q_{i,\rm fit}$ is given by Eq.~\eqref{eq:qifit}. Once the nested assignment is specified, all the quantities in Eqs.~\eqref{eq:Gfit} and~\eqref{eq:SFfit} are fixed by $m$, $b$ and $c$. The factors $N\alpha$ cancel because the Gini and Shannon measures depend on the relative member-wise distribution, not on its overall size.

The last two expressions require an important qualification. The fitted distribution $P_{\rm model}(k;b,c)$ gives the number of decisions at each consultation level, but it does not identify the members who were consulted. Different member assignments can therefore give the same fitted distribution and the same $I_1$, but different Gini and Shannon values. Equations~\eqref{eq:Fifit}--\eqref{eq:SFfit} therefore give model-based values under the stated nested assignment. They are not unique consequences of the fitted distribution.

For the two surveyed groups, Table~\ref{tab:fitted_indices} compares the values calculated from the fitted parameters with the values obtained from the data. We calculate the values of $I_2$ and $I_3$ in the data column from the observed aggregate consultation counts. The Gini and Shannon values in that column are from the illustrative nested member assignments described earlier. The fitted values of Gini and Shannon use the same nested assignment, but with the fitted distribution in place of the observed distribution.

\begin{table}[htbp]
\centering
\caption{Comparison of indices obtained from the data and from the fitted maximum-entropy distribution. The fitted Gini and Shannon values use the illustrative nested member assignment described in the text. The fitted member-wise counts are expected values under this assignment.}
\label{tab:fitted_indices}
\small
\setlength{\tabcolsep}{5pt}
\begin{tabular}{lrrrr}
\toprule
& \multicolumn{2}{c}{Small group} & \multicolumn{2}{c}{Large group}\\
\cmidrule(lr){2-3}\cmidrule(lr){4-5}
Index & Data & Fitted & Data & Fitted\\
\midrule
$I_2=I_1$ & 0.4661 & 0.4747 & 0.4694 & 0.4815\\
$I_3(40\%)$ & 0.5956 & 0.6056 & 0.5960 & 0.5997\\
$I_3(50\%)$ & 0.4444 & 0.4543 & 0.4266 & 0.4386\\
$I_3(60\%)$ & 0.2578 & 0.2658 & 0.2966 & 0.3107\\
$I_3(70\%)$ & 0.1644 & 0.1717 & 0.1808 & 0.1960\\
$I_3(80\%)$ & 0.0933 & 0.1010 & 0.0979 & 0.1150\\
$G$ & 0.4303 & 0.4218 & 0.4200 & 0.4091\\
$S_F$ & 0.8933 & 0.8990 & 0.9325 & 0.9369\\
\bottomrule
\end{tabular}
\vspace{2mm}
\begin{minipage}{0.94\textwidth}
\footnotesize
\textit{Note:} The data values of $I_2$ and $I_3$ use the observed aggregate decision-level counts. The data values of $G$ and $S_F$ use the illustrative nested member assignments. The fitted values use the fitted $b$ and $c$ values reported for the two groups. For $G$ and $S_F$, the same nested assignment is used with the fitted distribution. Therefore, the comparison is a consistency check, not an independent validation of member-wise inequality.
\end{minipage}
\end{table}

The fitted values are reasonably close to the corresponding data values. For the two groups, the fitted values give the same ordering of the Gini and Shannon measures. The large group has the lower Gini coefficient and the higher Shannon participation index. This is consistent with a somewhat more even distribution of consultation among members in the large-group example. This agreement provides a useful consistency check. It does not mean that the fitted distribution uniquely determines member-wise inequality or evenness.

Table~\ref{tab:summary_indices} summarises the main consultation and participation measures for the two surveyed groups.

\begin{table}[htbp]
\centering
\caption{Summary of consultation and participation measures for the two surveyed groups.}
\label{tab:summary_indices}
\small
\setlength{\tabcolsep}{4pt}
\begin{tabularx}{\textwidth}{@{}>{\raggedright\arraybackslash}p{0.19\textwidth} >{\raggedright\arraybackslash}p{0.25\textwidth} >{\raggedright\arraybackslash}X >{\centering\arraybackslash}p{0.09\textwidth} >{\centering\arraybackslash}p{0.09\textwidth}@{}}
\toprule
Measure & Definition & Meaning / significance & Small group & Large group \\
\midrule
Decision-side consultation index $I_1$ & $I_1=\frac{1}{mN}\sum_r k_r$ & Overall breadth of consultation across decisions. & 0.4661 & 0.4694 \\
Member-side consultation index $I_2$ & $I_2=\frac{1}{m}\sum_i f_i$ & Overall breadth of consultation viewed from the members' side. It is mathematically equal to $I_1$. & 0.4661 & 0.4694 \\
Threshold index $I_3(40\%)$ & Fraction of decisions with at least $40\%$ of members consulted. & Shows how often a chosen minimum level of consultation is reached. & 0.5956 & 0.5960 \\
Threshold index $I_3(50\%)$ & Fraction of decisions with at least $50\%$ of members consulted. & Same measure at a higher consultation threshold. & 0.4444 & 0.4266 \\
Threshold index $I_3(60\%)$ & Fraction of decisions with at least $60\%$ of members consulted. & Same measure at a higher consultation threshold. & 0.2578 & 0.2966 \\
Threshold index $I_3(70\%)$ & Fraction of decisions with at least $70\%$ of members consulted. & Shows how often broad consultation reaches a relatively high threshold. & 0.1644 & 0.1808 \\
Threshold index $I_3(80\%)$ & Fraction of decisions with at least $80\%$ of members consulted. & Shows how often very broad consultation is reached. & 0.0933 & 0.0979 \\
Gini coefficient $G$ & Gini coefficient of the member-wise consultation counts $F_i$. & Measures inequality in consultation among members. Higher values mean greater inequality. & 0.4303 & 0.4200 \\
Normalised Shannon participation index $S_F$ & $S_F=-\sum_i q_i\ln q_i/\ln m$, with $q_i=F_i/(\sum_j F_j)$. & Measures evenness of consultation among members. Higher values mean greater evenness. & 0.8933 & 0.9325 \\
\bottomrule
\end{tabularx}
\vspace{2mm}
\begin{minipage}{0.96\textwidth}
\footnotesize
\textit{Note:} The values of $I_1$, $I_2$ and $I_3(x)$ are based on the observed aggregate decision-level data for the two surveyed groups. The values of $G$ and $S_F$ are calculated from the illustrative member-wise assignments used to demonstrate the method. The aggregate $n_k$ values do not identify which members were consulted and therefore do not determine $G$ or $S_F$ uniquely.
\end{minipage}
\end{table}

\section{Simulations}

\subsection{Preferential consultation simulation}\label{sec:simulation}

\subsubsection{Simulation scheme}
The survey responses suggest that the coordinator may prefer certain members. This observation is not an estimate of a numerical preference parameter. In the simulation, we represent it using a simple model. A fixed group is taken as the \emph{preferred pool}. In any particular decision, only some members of this pool may be consulted. Thus, the pool remains fixed, but the members actually consulted can change from one decision to another. Other members can also be consulted when needed.

The survey indicates that some members are preferred. The simulation examines what may happen when this preference is repeated. These members may then be consulted more often even though the total number consulted remains unchanged.

The simulation does not aim to reproduce all the behaviour of a real coordinator. It examines, under controlled conditions, whether repeated preference can lead to unequal consultation among members.

We use a preferred pool of $m_p=6$ members for both group examples. This corresponds to $m_p/m=27.3\%$ for the small group and $m_p/m=5.7\%$ for the large group. The choice $m_p=6$ is illustrative and was not estimated from the survey.

The observed consultation count $k_r$ is kept fixed for each decision, so the simulation does not change the number of consultees. If $k_r\leq m_p$, the first $k_r$ members of the preferred pool are selected. If $k_r>m_p$, all $m_p$ preferred members are selected first, and the remaining $k_r-m_p$ members are selected from outside the pool. For these additional members, two choices are possible. With probability $p$, they follow a fixed preference order; with probability $1-p$, they are selected at random from the remaining members. Thus, $p$ determines how often the fixed preference order is used. At $p=0$ the additional members are selected randomly, while at $p=1$ they follow the fixed order. Intermediate values combine the two choices.

This construction creates an \emph{active subset of the preferred pool}. The members selected from the pool can change with $k_r$, although the larger preferred pool remains fixed. As $p$ increases, the simulation uses the fixed preference order more often. This can make consultation more concentrated among the preferred members.

For each simulated set of decisions, we calculate the consultation count for each member as
\begin{equation}
F_i^{\rm sim}=\sum_r x_{ir}^{\rm sim},
\end{equation}
where $x_{ir}^{\rm sim}=1$ when member $i$ is selected for consultation in decision $r$ and is zero otherwise. We then calculate $I_2$, $G$ and $S_F$ from these counts. For each value of $p$, we perform 2000 independent simulation runs. Table~\ref{tab:simulation} gives the mean values from these runs.

A useful check follows directly from this construction. Exactly the observed $k_r$ members are selected in every simulated decision. Hence,
\begin{equation}
\sum_i F_i^{\rm sim}=\sum_r k_r.
\end{equation}
Hence,
\begin{equation}
I_{2,\rm sim}=\frac{1}{mN}\sum_iF_i^{\rm sim}
=\frac{1}{mN}\sum_rk_r=I_1.
\end{equation}
Thus, the simulation also satisfies the analytical identity $I_2=I_1$. The value of $I_2$ is unchanged by which members are selected, as long as the observed $k_r$ values remain fixed.

Table~\ref{tab:simulation} gives the results for the two group examples. The value of $I_2$ stays fixed as $p$ changes. The Gini coefficient, however, increases as the fixed preference order is used more often, while the normalised Shannon participation index decreases. Even at $p=0$, the fixed preferred pool makes consultation less even than fully random selection. With increasing $p$, the additional consultees are more often chosen according to the same order, and consultation becomes more concentrated.

\begin{table}[htbp]
\centering
\caption{Simulation of preferential consultation for the two group examples. The observed $k_r$ values are kept fixed, and a preferred pool of $m_p=6$ members is used. Each entry is the mean over 2000 simulation runs.}
\label{tab:simulation}
\setlength{\tabcolsep}{5pt}
\begin{tabular}{c ccc ccc}
\toprule
$p$ & \multicolumn{3}{c}{Small group} & \multicolumn{3}{c}{Large group} \\
\cmidrule(lr){2-4}\cmidrule(lr){5-7}
 & $I_2$ & $G$ & $S_F$ & $I_2$ & $G$ & $S_F$ \\
\midrule
0    & 0.4661 & 0.2931 & 0.9460 & 0.4694 & 0.0777 & 0.9937 \\
0.25 & 0.4661 & 0.3179 & 0.9430 & 0.4694 & 0.1537 & 0.9904 \\
0.50 & 0.4661 & 0.3545 & 0.9341 & 0.4694 & 0.2420 & 0.9803 \\
0.75 & 0.4661 & 0.3921 & 0.9186 & 0.4694 & 0.3309 & 0.9623 \\
1.00 & 0.4661 & 0.4303 & 0.8933 & 0.4694 & 0.4200 & 0.9325 \\
\bottomrule
\end{tabular}
\vspace{1mm}
\begin{minipage}{0.92\textwidth}
\footnotesize
\textit{Note:} $m_p=6$ is used in both examples. The parameter $p$ controls the choice of the additional members outside the preferred pool. At $p=0$ these additional members are selected at random; at $p=1$ they follow the fixed preference order. The simulation keeps the observed $k_r$ values unchanged. The member-wise values used elsewhere in the paper are illustrative assignments. The simulation is a demonstration of a possible mechanism of preferential consultation, not an empirical estimate of $p$ for either surveyed group.
\end{minipage}
\end{table}

\begin{figure}[htbp]
\centering
\includegraphics[width=0.88\textwidth]{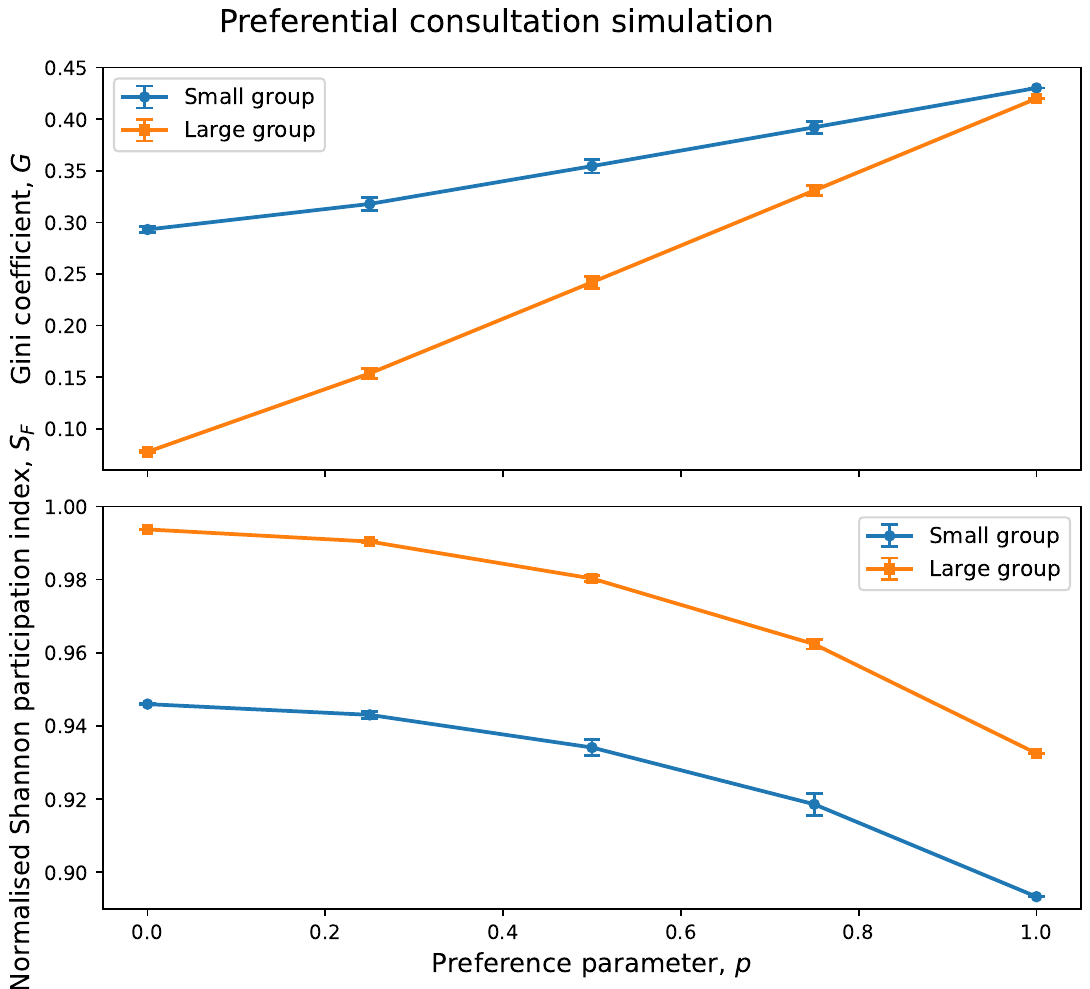}
\caption{Preferential-consultation simulation with the observed number of consultees fixed in every decision and a preferred pool of six members. The upper panel shows how the Gini coefficient increases as reliance on the fixed preference order for additional consultees increases. The lower panel shows the corresponding decrease in the normalised Shannon participation index. Error bars show one standard deviation across the 2000 runs. For some points, the standard deviations are very small, so the error bars appear very small at this scale. The value of $I_2$ is unchanged and is not plotted.}
\label{fig:simulation}
\end{figure}

\subsubsection{Simulation results}
At $p=0$, the preferred pool is used and the additional members are selected randomly. As $p$ increases, the fixed preference order is used more often. The Gini coefficient increases and the normalised Shannon participation index decreases. The value of $I_2$ remains unchanged because the observed $k_r$ values are kept fixed.

The corresponding results are shown in Figure~\ref{fig:simulation}.

\textbf{Survey feature and simulation result.} The survey provides the observation that the coordinator may prefer a particular group. The simulation shows a consequence of such repeated preference: consultation can become more concentrated among preferred members even though the overall breadth of consultation stays the same.

The main point is: \textbf{the same number of people can be consulted in each decision, but the same people can receive more of the consultation.} Therefore, the overall consultation level does not by itself tell us how evenly participation is shared among members.

\subsection{Fixed-core consultation simulation}\label{sec:coreperiphery}

The first simulation represents repeated preference for a larger preferred pool. We now consider a different case in which a fixed group of members is consulted in every decision. Additional members are then selected from the remaining group.

\subsubsection{Simulation scheme}
A fixed group forms the consultation core, and every member of this core is consulted in every decision. The other members form the peripheral group. For each decision, a specified number of peripheral members is selected at random. Thus, the core remains fixed, while the additional participants can change from one decision to another.

This simulation asks whether a permanent core can make consultation less even while the overall consultation level is kept fixed in expectation.

The core--periphery simulation matches the mean consultation level, but it does not preserve the full empirical sequence of consultation counts $k_r$. Its member-wise Gini values therefore need not lie within the bounds obtained when the empirical consultation-count distribution is held fixed.

This differs from the first simulation, where the preferred pool is fixed but only some of its members may be selected in a particular decision. Here, every member of the fixed core is consulted in every decision.

Let $s$ be the nominal fraction of members assigned to the core. Since the number of members must be an integer, the actual core size in a finite group is
\begin{equation}
c_{\rm core}=\operatorname{round}(sm),
\label{eq:core_size}
\end{equation}
where $c_{\rm core}$ is the number of core members. Here $c$ is the exponential parameter of the maximum-entropy model and should not be confused with $c_{\rm core}$. The remaining number of members is $m-c_{\rm core}$. Thus, in the finite simulation we use $c_{\rm core}$, while $s$ continues to denote the nominal core fraction.

Let $\rho$ be the target fraction of peripheral members consulted in a decision. For the finite-group simulation, the expected number of peripheral consultees is
\begin{equation}
q_{\rm per}=\rho(m-c_{\rm core}),
\label{eq:peripheral_q}
\end{equation}
Here $q_{\rm per}$ need not be an integer. In the finite simulation, however, the total number of consultees must be an integer. We therefore use two adjacent integer values of $k$ and choose their probabilities so that the expected value of $I_1$ matches the observed value. Peripheral members are selected uniformly without replacement for each decision. For a realised value of $k$, the number of peripheral consultees is $k-c_{\rm core}$. The quantity $q_{\rm per}$ in Eq.~\eqref{eq:peripheral_q} is the target mean number of peripheral consultees used in the finite-group calibration.
In the continuous calibration, the expected total number of consulted members is $c_{\rm core}+q_{\rm per}$. The corresponding mean consultation index is
\begin{equation}
I_{1,\rm cont}=\frac{c_{\rm core}+q_{\rm per}}{m}.
\label{eq:coreperiphery_I1}
\end{equation}

In the continuous idealisation, where $sm$ and $\rho(m-sm)$ can be treated without integer rounding, the mean consultation index is
\begin{equation}
I_1(s,\rho)=s+\rho(1-s).
\label{eq:coreperiphery_I1_cont}
\end{equation}
The first term comes from the permanent core, and the second term comes from the peripheral members.

The model also gives a simple expression for member-wise consultation inequality in the large-sample limit. A core member is consulted in every decision, whereas a peripheral member is consulted with frequency $\rho$. The Gini coefficient for these two consultation frequencies is
\begin{equation}
G=\frac{s(1-s)(1-\rho)}{I_1},
\label{eq:coreperiphery_G}
\end{equation}
provided $I_1>0$. This follows from the pairwise definition of the Gini coefficient. Only pairs containing one core member and one peripheral member have a nonzero difference in consultation frequency. If $I_1$ is fixed, Eq.~\eqref{eq:coreperiphery_I1_cont} gives
\begin{equation*}
\rho=\frac{I_1-s}{1-s}.
\end{equation*}
Substituting this into Eq.~\eqref{eq:coreperiphery_G}, we obtain
\begin{equation*}
1-\rho=1-\frac{I_1-s}{1-s}=\frac{1-I_1}{1-s},
\end{equation*}
and hence
\begin{equation*}
G=\frac{s(1-s)}{I_1}\frac{1-I_1}{1-s}=\frac{s(1-I_1)}{I_1}.
\end{equation*}
Thus, in this continuous idealisation, $G$ increases with the core fraction over the feasible range $0\leq s\leq I_1<1$, when $I_1$ is fixed. The finite simulations below provide a numerical check of this result.

For a finite group with $c_{\rm core}$ core members, the expected member-wise consultation frequencies are
\begin{equation}
f_{\rm core}=1,\qquad f_{\rm per}=\rho.
\label{eq:coreperiphery_freq}
\end{equation}
The corresponding normalised Shannon index is
\begin{equation}
S_F=-\frac{c_{\rm core} \pi_{\rm core}\ln \pi_{\rm core}+(m-c_{\rm core})\pi_{\rm per}\ln \pi_{\rm per}}{\ln m},
\label{eq:coreperiphery_SF}
\end{equation}
where
\begin{equation}
\pi_{\rm core}=\frac{1}{mI_1},\qquad \pi_{\rm per}=\frac{\rho}{mI_1}.
\label{eq:coreperiphery_ab}
\end{equation}
Terms with zero probability are taken as zero. These expressions provide analytical reference values. The simulation values reported below are calculated directly from the simulated member-wise consultation counts.

We checked numerical stability by running both simulations with 500, 1000 and 2000 independent repetitions for each parameter value. The mean estimates changed only slightly as the number of repetitions increased. For Simulation~1, the largest change from 500 to 1000 runs was $4.1\times10^{-4}$ for $G$, and the largest change from 1000 to 2000 runs was $3.8\times10^{-4}$. For Simulation~2, the corresponding changes were $5.4\times10^{-4}$ and $1.74\times10^{-4}$. The changes in $S_F$ were smaller. We therefore use 2000 repetitions for the reported results. This check concerns numerical stability; it is not a formal proof of convergence.

\subsubsection{Effect of core size}

\begin{figure}[htbp]
\centering
\includegraphics[width=0.82\textwidth,height=0.70\textheight,keepaspectratio]{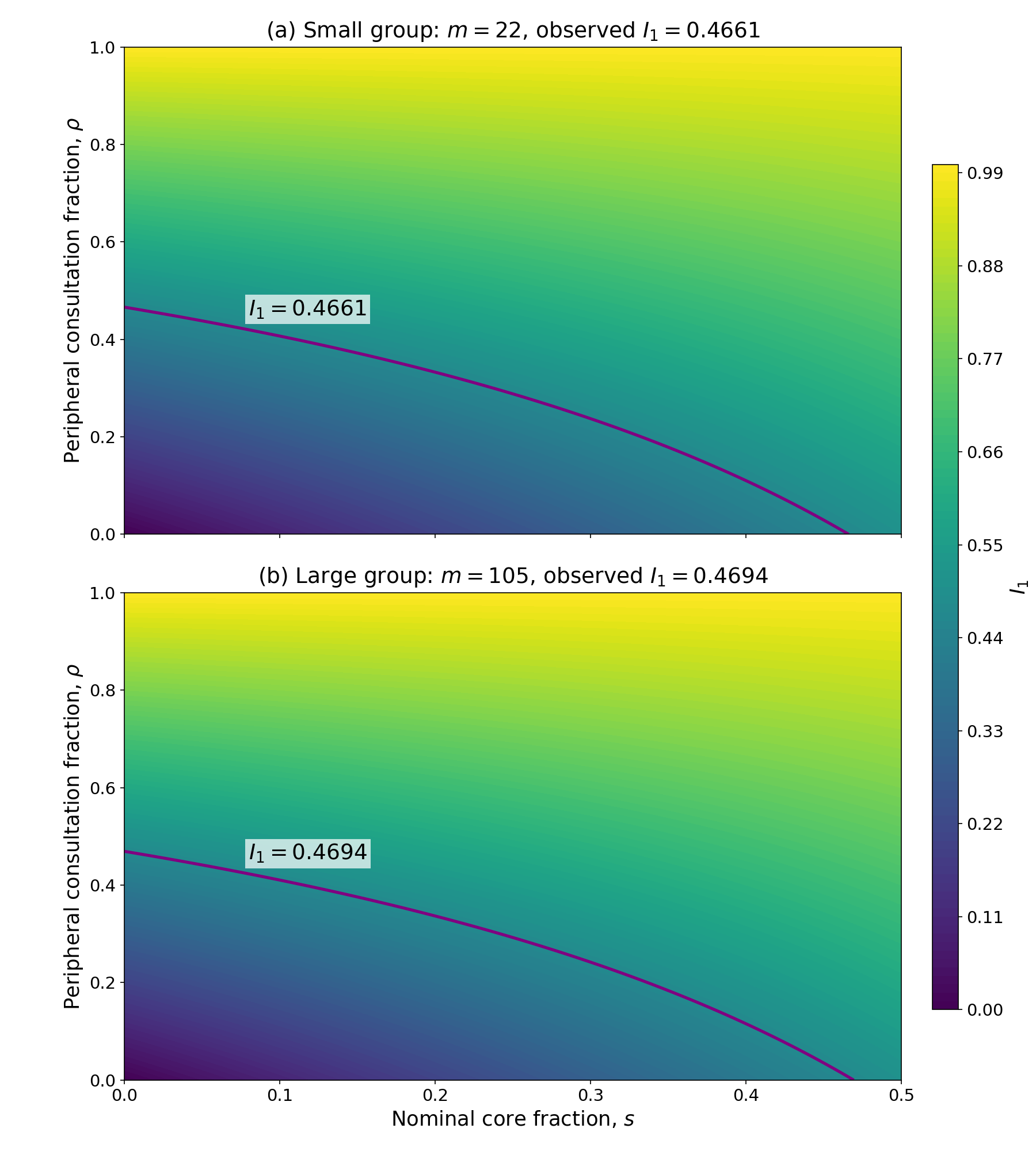}
\caption{Parameter space of the core--periphery model. The colour scale shows the consultation index $I_1(s,\rho)=s+\rho(1-s)$. The two marked contours correspond to the observed consultation levels of the small and large surveyed groups. Moving along either contour keeps the overall consultation level unchanged. It changes the balance between the permanent core and the peripheral group.}
\label{fig:scheme2I1}
\end{figure}

We first examine the model over $0\leq s\leq0.5$ and $0\leq\rho\leq1$. Figure~\ref{fig:scheme2I1} shows the resulting consultation index and marks the two observed levels, $I_1=0.4661$ and $I_1=0.4694$. A constant-$I_1$ contour shows that the same overall consultation level can be obtained for different core fractions. Along such a contour, a larger core requires a smaller peripheral consultation fraction.

This is the main reason for introducing the core--periphery model. The value of $I_1$ gives the overall breadth of consultation, but it does not show whether the same members are consulted repeatedly. The parameters $s$ and $\rho$ separate these two aspects.

\subsubsection{Simulation at the observed consultation level}

We next examine this effect numerically for the two surveyed group sizes. The purpose is not to fit the simulation to the survey data. Instead, the observed average consultation level is used as the target while the core fraction is varied. The survey data do not identify a particular core size.

For each nominal core fraction, the actual core size was obtained from Eq.~\eqref{eq:core_size}. Peripheral members were selected uniformly without replacement. As in Simulation~1, we used 2000 independent runs for each parameter value.

Two finite-size points are important here. First, $sm$ need not be an integer, so the rounded core size is the actual value used in the simulation. Second, the observed mean consultation count need not be an integer. We therefore use the two adjacent integer values of $k$ so that their weighted mean equals the observed value. For the small group, we use $k=10$ and $k=11$ with probabilities $0.746667$ and $0.253333$, giving an expected value of $10.253333$. For the large group, we use $k=49$ and $k=50$ with probabilities $0.7146893$ and $0.2853107$, giving an expected value of $49.2853107$. In general, if the target mean is $\mu$, the two adjacent integers are $k_1=\lfloor\mu\rfloor$ and $k_2=k_1+1$, with probabilities $k_2-\mu$ and $\mu-k_1$. These probabilities sum to one and give an expected value of $k$ equal to $\mu$. Hence, the expected value of $I_1$ equals the observed value for each group.

In each run, the core members remain fixed for all decisions. The peripheral members are sampled independently and uniformly without replacement for each decision, subject to the required number of peripheral consultees. After all decisions are generated, the member-wise counts $F_i$ are calculated and used to obtain $I_2$, $G$ and $S_F$. The same procedure is repeated for all core fractions.

The results are shown in Table~\ref{tab:coreperiphery_sim}. The mean values of $I_1$ are close to the target values. For the cases with a nonzero permanent core, $G$ increases and $S_F$ decreases as the core fraction increases. In the small group, the first two nominal values of $s$ both give zero core members after rounding, so they represent the same finite model condition.

As a simple control, we also ran the fixed-$k$ version of the model. We used $k=10$ for the small group and $k=49$ for the large group, the nearest whole numbers to their observed mean consultation counts. In this control, $I_1$ is fixed exactly within each run. For increasing nonzero core sizes, we again obtained an increase in $G$ and a decrease in $S_F$. This control is not used for the main comparison because its consultation levels differ slightly from the observed values. It is included to check that the main result is not caused only by the small random variation in $k$ used in the empirical-$I_1$ calibration.

\begin{table}[htbp]
\centering
\caption{Core--periphery simulation calibrated to the observed mean consultation level. Each entry is the mean over 2000 simulation runs. The core size is $c_{\rm core}=\operatorname{round}(sm)$. The peripheral fraction $\rho$ is chosen so that the expected consultation level equals the observed mean.}
\label{tab:coreperiphery_sim}
\small
\setlength{\tabcolsep}{4pt}
\begin{tabular}{c rr rr rr}
\toprule
Nominal $s$ & \multicolumn{3}{c}{Small group} & \multicolumn{3}{c}{Large group}\\
\cmidrule(lr){2-4}\cmidrule(lr){5-7}
 & $c_{\rm core}$ & $\rho$ & $G$ & $c_{\rm core}$ & $\rho$ & $G$\\
\midrule
0.01 & 0 & 0.4661 & 0.0394 & 1 & 0.4643 & 0.0287\\
0.02 & 0 & 0.4661 & 0.0394 & 2 & 0.4591 & 0.0392\\
0.05 & 1 & 0.4406 & 0.0879 & 5 & 0.4429 & 0.0703\\
0.10 & 2 & 0.4127 & 0.1361 & 10 & 0.4135 & 0.1224\\
0.20 & 4 & 0.3474 & 0.2331 & 21 & 0.3367 & 0.2372\\
0.30 & 7 & 0.2169 & 0.3794 & 32 & 0.2368 & 0.3520\\
0.40 & 9 & 0.0964 & 0.4766 & 42 & 0.1156 & 0.4564\\
\bottomrule
\end{tabular}
\vspace{1mm}
\begin{minipage}{0.94\textwidth}
\footnotesize
\textit{Note:} The small-group target is $I_1\approx0.4661$ and the large-group target is $I_1\approx0.4694$. The simulated mean values of $I_1$ differ from the targets only by small simulation fluctuations. For the small group, $s=0.01$ and $s=0.02$ give $c_{\rm core}=0$ after rounding. These two cases therefore do not contain a nonzero permanent core.
\end{minipage}
\end{table}

\begin{figure}[htbp]
\centering
\includegraphics[width=0.88\textwidth]{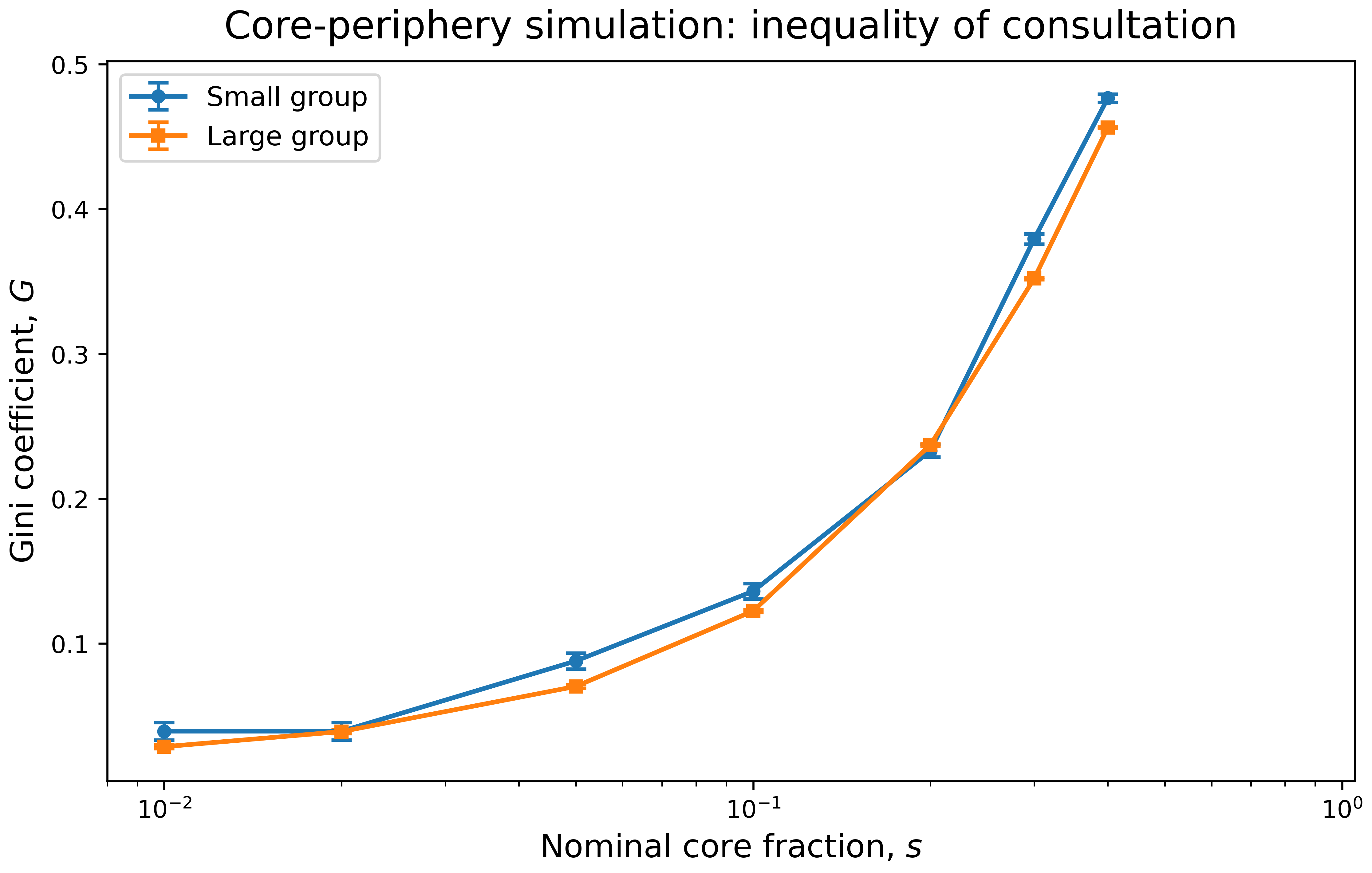}
\caption{Gini coefficient from the core--periphery simulation at the observed consultation level. The points are means over 2000 simulation runs, with error bars showing one standard deviation. The overall consultation level is calibrated to the observed mean for each group. As the permanent core becomes larger, consultation becomes more unequal among members. For some points, the standard deviations are very small, so the error bars appear very small at this scale.}
\label{fig:scheme2G}
\end{figure}

Figure~\ref{fig:scheme2G} shows the Gini coefficient for the simulation calibrated to the observed consultation level. The increase does not come from a change in the overall consultation level. The expected value of $I_1$ remains at the observed value. The change comes from how consultation is shared among members.

\begin{figure}[htbp]
\centering
\includegraphics[width=0.88\textwidth]{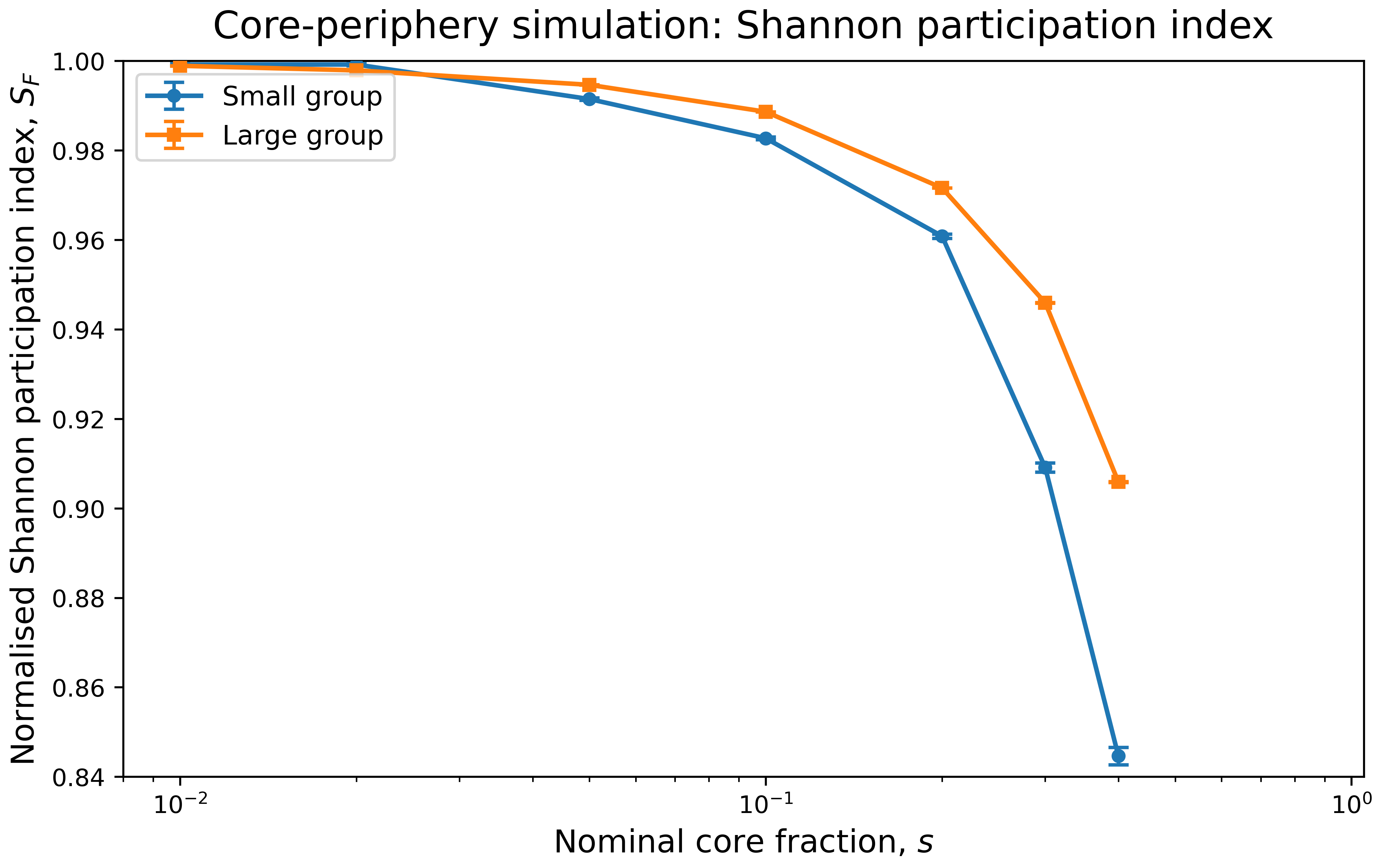}
\caption{Normalised Shannon participation index from the core--periphery simulation at the observed consultation level. The points are means over 2000 simulation runs, with error bars showing one standard deviation. The decrease shows that consultation becomes less evenly distributed among members as the permanent core becomes larger. For some points, the standard deviations are very small, so the error bars appear very small at this scale.}
\label{fig:scheme2SF}
\end{figure}

Figure~\ref{fig:scheme2SF} shows the corresponding normalised Shannon participation index. It decreases as the core becomes larger. Thus, the Gini and Shannon measures describe the same change in opposite directions: a larger permanent core means greater inequality and less evenness in consultation.

The model can be understood simply. Suppose two groups have nearly the same value of $I_1$. In one group, the members consulted in one decision may be quite different from those consulted in the next. In another group, a smaller set of members may be consulted repeatedly, while the remaining consultation is shared among the other members. The two groups can then have similar values of $I_1$ but different values of $G$ and $S_F$.

The simulation does not show that the surveyed groups actually had the core--periphery structure used here. The survey data do not contain the individual consultation histories needed to identify such a structure. Instead, the simulation gives a controlled example of how a permanent core can change the distribution of consultation. It shows that the overall consultation level can remain fixed in expectation while consultation becomes less evenly shared among members.

\subsubsection{Exploratory variation of the core--periphery model}

We also considered a small variation of the core--periphery model as an exploratory check. Here, the additional members are not chosen completely at random. Instead, they are given a fixed order of preference. The number of additional members can vary from one decision to another, up to a chosen maximum fraction of the members outside the core. The first members in the fixed order are then selected. For this exploratory calculation, the number of additional members is chosen uniformly from zero to the chosen maximum. Thus, members outside the core do not all have the same chance of being consulted.

This calculation asks whether this simple preference pattern can produce values of $I_1$, $G$ and $S_F$ reasonably close to those from the synthetic member-wise assignments in Tables~\ref{tab:table3} and~\ref{tab:table4}. It is not a fit to the survey data. We varied the nominal core fraction $s$ from 0.05 to 0.40 and the maximum additional fraction from 0.10 to 1.00. For each combination, we used 500 simulation runs. Some combinations gave all three quantities reasonably close to the corresponding illustrative values. Here, closeness means small unweighted differences in $I_1$, $G$ and $S_F$; it does not mean a statistical fit. The parameter values should not be interpreted as estimates of the actual size of a preferred group. The survey data do not provide the individual consultation histories needed for such an estimate.

This variation is useful because the core--periphery model treats all members outside the core in the same way. In an actual group, members outside a preferred group may also differ in their relationship with the coordinator. The exploratory calculation shows only how such an additional preference pattern can affect the member-wise measures. It is not presented as a third simulation or as a fitted model of the surveyed groups. The exploratory result file is included in the submission package along with this article.

\subsection{Comparison of the two simulation schemes}

The two simulations describe two different ways in which consultation can become concentrated among members.

\begin{center}
\begin{minipage}{0.92\textwidth}
\textbf{Simulation 1.} A fixed group of members is preferred for consultation, but not all of them are consulted in every decision. The number of people consulted in each decision is kept equal to its observed value in the simulation process. When more members are needed than are available in the preferred group, the remaining members are selected from outside this group. A parameter in the simulation model controls whether this selection is mainly based on the coordinator's preference or made more randomly.

\medskip
\textbf{Simulation 2.} A fixed group of members is consulted in every decision. Additional members are selected from the remaining group, and their number can change from one decision to another. Thus, some members form a permanent core of consultation, while others are consulted only occasionally. The size of the permanent group and the number of additional members are controlled by parameters defined for this simulation. The overall number of consultations, summed over all decisions, is kept at the observed level on average.

\medskip
\textbf{Common result.} The two simulations show that the same overall level of consultation can be shared among members in different ways. In particular, consultation can become concentrated among some members even when the overall amount of consultation remains the same.
\end{minipage}
\end{center}

The first simulation is motivated by the survey observation that the coordinator may prefer a particular group of members. It shows one possible effect of repeated preference. The second simulation is not presented as a description of the surveyed groups. It is a controlled model experiment showing one possible effect of a permanent consultation core.

The distinction matters. In the first simulation, the number of people consulted in each decision is kept the same, while the identities of those people can change. This allows us to examine the effect of repeated preference without changing the consultation level of any decision. In the second simulation, a fixed group is consulted in every decision, while the remaining consultation is shared among other members. The overall consultation level is kept the same on average.

Taken together, the simulations show that \textbf{the total amount of consultation and its distribution among members are different properties of a group decision process.} A single measure of overall breadth cannot describe both.

The model also has a clear limitation. In the fixed-size version, the number of additional members consulted is fixed within a given simulation condition. The version used for the reported results allows two nearby consultation counts. This keeps the overall consultation level at the observed value on average. The model does not reproduce the full range of consultation levels seen in the survey. It is used here to study how consultation can be distributed unequally among members at a controlled overall level. A future extension could allow the number of additional members to vary and could then be compared directly with the full survey distribution.

\section{Discussion}

\subsection{Interpretation of the index}

The proposed index $I_1$ has a direct interpretation. It is the average fraction of available members who are consulted per decision. In the small-group aggregate, $I_1\approx0.4661$, while in the large-group aggregate $I_1\approx0.4694$. Thus, the two case groups have similar overall consultation breadth. This result follows directly from the decision records and does not depend on the probability model.

The distribution of consultation levels gives information that is not available from the average alone. In the small-group example, the average number of members consulted is $10.253333$, while $P(k)$ is largest at $k=8$. For the large group, the corresponding values are $49.2853107$ and $42$. The mean and mode need not be equal. Similarly, the cumulative quantity $P_{\geq}(k)$ gives threshold information that is not contained in $I_1$. For example, $P_{\geq}(12)=0.3778$ in the small-group case, so 37.78\% of the decisions involve more than half of the 22 members.

The member-side index gives the same overall measure by using the fraction of decisions in which each member is consulted. Section~4 proves that $I_2=I_1$, since both indices count the same consultation events. When the identities of consulted members are known, the member-wise frequencies $F_i$ provide additional information. The Gini coefficient, Lorenz curve and Shannon participation index then describe how evenly consultation is distributed among members.

The probability model describes the distribution of consultation counts. The quantity $b/c$ gives the location of the maximum of the continuous kernel. The discrete mode is found by comparing the allowed integer values of $k$. Equation~\eqref{eq:cov} also shows that, when $m$ and $c$ are fixed, increasing $k_0=b/c$ increases the mean number of consulted members and hence $I_1$. Since $I_1$ depends on both $k_0$ and $c$, $k_0$ alone should not be used to compare groups with different fitted values of $c$.

The nested assignments in Tables~\ref{tab:table3} and~\ref{tab:table4} illustrate how member-wise calculations can be made from aggregate data. The aggregate $n_k$ values do not identify which members were consulted, so the individual $F_i$ values cannot be determined uniquely. Other assignments consistent with the same decision-level counts can give different member-wise distributions, but the total remains fixed by the counting identity. Thus, $I_2=I_1$ does not depend on the particular illustrative assignment.

Equation~\eqref{eq:I1decision} also gives a useful sensitivity bound. If the consultation count for each decision changes by at most one member, the change in $I_1$ is at most $1/m$. More generally, if $L_{\Delta}=\sum_r|\Delta k_r|$ is the total absolute change in the recorded counts, then
\begin{equation*}
|\Delta I_1|\leq\frac{L_{\Delta}}{mN}.
\end{equation*}
This gives a simple way to assess the effect of a small number of uncertain consultation events.

\subsection{Possible factors affecting consultation}

The interviews suggest several possible reasons why some members may be consulted less often. A member may be unwilling or unable to give an opinion, or may have limited interest, knowledge or experience relevant to a particular matter. Personal circumstances may also affect participation. Some respondents mentioned differences between members and the coordinator, as well as repeated non-response to requests for views. We could not independently check these reports because of the confidentiality conditions, so we treat them only as possible explanations.

Some respondents said that coordinators relied more often on a smaller set of members for important decisions. This was reported in both small and large groups, although the extent varied. Such reports are consistent with the possibility that consultation can be broad while decision-making influence remains concentrated among fewer members. The consultation records alone cannot establish the reason for non-consultation or the amount of influence given to a consulted member.

The absence of zero-consultation decisions in the small-group aggregate may also have more than one explanation, such as closer interaction among members or greater dependence on member views. The available data do not distinguish between these possibilities. Thus, $n_0$ and $I_1$ describe the observed consultation pattern, while the reasons behind that pattern require additional information.

The main limitation of the index is that it measures whether a member is consulted, not how much influence that member has on the final decision. It does not measure the quality of discussion, freedom to disagree, equality of influence, transparency or whether the final decision follows the majority view. Therefore, $I_1$ should be understood as a measure of participation through consultation and as one aspect of democratic decision making, not as a complete measure of democracy.

The maximum-entropy form is also a descriptive model. Its role is to describe the observed consultation-count distribution under the stated constraints; it is not a theory of how coordinators choose members for consultation. The definition of $I_1$ does not depend on this modelling choice.

\subsection{Information requirements for practical application}

For a real group, the decisions and the period over which the index will be measured should first be defined. A consistent rule should then be used to decide when a member is regarded as consulted. If the records give only $k_r$ for each decision, we can calculate $I_1$ and $I_3(x)$. If they also identify the members consulted, we can calculate $F_i$, the Gini coefficient, the Lorenz curve, the Shannon participation index and the member--decision network. Alternatively, a member-side survey can record the fraction of decisions in which each member's opinion was sought. The average of these fractions gives $I_2$ and equals $I_1$ when the records are complete.

The choice of the observation period can affect the result. A short period may contain too few decisions, while a longer period may include changes in membership or decision procedures. Here we assume that the set of $m$ members remains fixed. If membership changes, the data can be divided into suitable periods or adjusted so that the denominator $mN$ has a clear meaning. The calculation is simple, but its value depends on a clear and consistent definition of consultation and reliable records.

\section{Conclusion}

This study presents a way to measure how widely non-coordinator members are consulted in a group. The main index $I_1$ is the average fraction of available members consulted per decision. The member-side index $I_2$ gives the average fraction of decisions in which a member's opinion is sought. We have shown analytically that $I_2=I_1$. The threshold index $I_3(x)$ gives the fraction of decisions that involve at least a chosen percentage ($x$) of members. Together, these indices provide different views of the same consultation process.

The mathematical framework also gives useful relations among these measures. A complete threshold function determines $I_1$, while a single threshold gives bounds on $I_1$. The binomial calculation gives probabilities for future all-member consultation, assuming that the observed rate remains unchanged. The Poisson distribution provides an approximation for rare events, while the maximum-entropy model describes the observed distribution of consultation levels.

The two aggregate case groups have similar values of the main index: $I_1\approx0.4661$ for the small group and $I_1\approx0.4694$ for the large group. Their consultation-level distributions are not identical, however. The mean consultation level, the mode and the threshold probabilities provide additional information. When member identities are available, the Gini coefficient, Lorenz curve and Shannon participation index give a further description of how consultation is shared among members.

The simulations illustrate an important consequence of this distinction. The same overall consultation level can be produced by different patterns of member selection. Repeated preference for some members or a permanent consultation core can increase concentration among members without requiring a corresponding increase in the overall value of $I_1$. Thus, overall consultation breadth and the distribution of consultation among members are related but distinct properties of group decision making. The simulations are controlled demonstrations rather than fitted models of the surveyed groups.

The empirical study has some limitations. The detailed numerical analysis uses two groups selected from a survey of 102 groups; these cases are not statistical estimates of all 48 small groups or all 54 large groups. The member-wise examples are synthetic assignments constructed from the aggregates. The index $I_1$ measures consultation, not influence, and is one aspect of democratic participation within a group.

The framework can be applied when a group keeps a consistent record of decisions and the members whose opinions were sought. Using only the number of members consulted in each decision, $I_1$ and $I_3(x)$ can be obtained. With member identities, the member-wise measures and the member--decision network can also be studied. This makes the framework applicable to committees, academic departments, organisations, teams, boards, councils and other decision-making bodies. It can also be used to compare consultation patterns across different groups or across different periods in the same group, provided that the definition of consultation and the method of recording remain the same.

Larger datasets with complete member-level records would allow researchers to compare consultation breadth and member-wise inequality across different types of groups. They could also test whether the patterns suggested by the simulations occur in observed member--decision networks and examine how consultation relates to other aspects of participation and decision making. The framework provides a starting point for these studies by separating the overall breadth of consultation from the way consultation is distributed among members.

\section*{Competing interests}
The author declares no competing interests.

\section*{Data availability}

Tables~\ref{tab:table1} and~\ref{tab:table2} report the full aggregate consultation-count data used in the numerical analysis. These data come from documented decision records of two groups in the survey: one small group and one large group. The wider survey covered 102 groups, including 48 small groups and 54 large groups. The survey was carried out over four years through confidential, in-person interviews with coordinators and members, where we collected information on decisions taken by the groups during the five years preceding each interview.

The classification used here is based on the observed sample. The 48 small groups had $m\leq30$ and $n_0=0$, while the 54 large groups had $m>30$ and $n_0>0$. The boundary at $m=30$ applies to this sample and is not proposed as a general classification rule. The two groups examined in detail are not statistical estimates of all groups in their respective classes. The full survey provides the broader empirical context for these two cases.

The documented records included meeting details and minutes, consultation records, emails, written proposals and feedback, written opinions or assessments, online records, personal written records or evaluations, signed decisions or resolutions, decision proposals and other relevant documents. The published tables (Tables~1 and~2) contain only the aggregate consultation counts used in the analysis. Individual and organisational identities are not disclosed because the interviews were carried out under strict confidentiality conditions.

The aggregate data files \texttt{aggregate\_small.csv} and \texttt{aggregate\_large.csv} are included in the submission package. They contain the non-confidential aggregate consultation-count data listed in Tables~1 and~2. The member-wise calculations in Tables~\ref{tab:table3} and~\ref{tab:table4} use synthetic nested assignments constructed from these aggregates; they do not represent observed individual histories.

The simulation source files \texttt{simulation1.py} and \texttt{simulation2.py}, together with the exploratory result file \texttt{scheme2\_exploratory\_parameter\_sweep\_500\_results.csv}, are included in the submission package. These files contain the code for the simulation schemes and the results of the exploratory calculation described in the paper. Individual interview records and other records containing respondent or organisational identities are not included because they are confidential.
\bibliographystyle{unsrt}
\bibliography{references}

\end{document}